\documentclass[10pt,journal,compsoc]{IEEEtran}

\ifCLASSOPTIONcompsoc
  \usepackage[nocompress]{cite}
\else
  \usepackage{cite}
\fi

\usepackage{graphicx}
\usepackage{caption}

\newtheorem{qry}{Query}
\newtheorem{def1}{Definition}
\DeclareMathAlphabet\mathbfcal{OMS}{cmsy}{b}{n}

\begin{document}
\title{Privacy-Preserving Aggregate Queries for Optimal Location Selection}

\author{Emre~Yilmaz,
        Hakan~Ferhatosmanoglu,
				Erman~Ayday,
        and~Remzi~Can~Aksoy
\IEEEcompsocitemizethanks{\IEEEcompsocthanksitem E. Yilmaz and E. Ayday are with the Computer Engineering Department, Bilkent University, Ankara 06800, Turkey. E-mail: [emre.yilmaz, erman]@cs.bilkent.edu.tr
\IEEEcompsocthanksitem H. Ferhatosmanoglu is with the Department of Computer Science, University of Warwick, Coventry, CV4 7AL, UK, and the Computer Engineering Department, Bilkent University, Ankara 06800, Turkey. E-mail: hakan.f@warwick.ac.uk
\IEEEcompsocthanksitem R. Aksoy was with Bilkent University, Ankara 06800, Turkey. He is now with the University of Michigan, Ann Arbor, MI 48109. E-mail: remzican@umich.edu.
\IEEEcompsocthanksitem This work is supported in part by Turk Telekom.}
\thanks{}}

\markboth{IEEE TRANSACTIONS ON DEPENDABLE AND SECURE COMPUTING}%
{Shell \MakeLowercase{\textit{et al.}}: Bare Demo of IEEEtran.cls for Computer Society Journals}

\IEEEtitleabstractindextext{%
\begin{abstract}
Today, vast amounts of location data are collected by various service providers. These location data owners have a good idea of where their users are most of the time. Other businesses also want to use this information for location analytics, such as finding the optimal location for a new branch. However, location data owners cannot share their data with other businesses, mainly due to privacy and legal concerns. In this paper, we propose privacy-preserving solutions in which location-based queries can be answered by data owners without sharing their data with other businesses and without accessing sensitive information such as the customer list of the businesses that send the query. We utilize a partially homomorphic cryptosystem as the building block of the proposed protocols. We prove the security of the protocols in semi-honest threat model. We also explain how to achieve differential privacy in the proposed protocols and discuss its impact on utility. We evaluate the performance of the protocols with real and synthetic datasets and show that the proposed solutions are highly practical. The proposed solutions will facilitate an effective sharing of sensitive data between entities and joint analytics in a wide range of applications without violating their customers' privacy.

\end{abstract}

\begin{IEEEkeywords}
Privacy, Data encryption, Security, integrity, and protection, Query processing, Algorithm/protocol design and analysis.
\end{IEEEkeywords}}

\maketitle



\IEEEraisesectionheading{\section{Introduction}\label{sec:introduction}}

\IEEEPARstart{U}{nderstanding} the whereabouts of current and potential customers can provide valuable insights for location-based services, facility location, and competitive business decisions. Increasing amounts of location data from mobile services, applications, and network operators have introduced exciting opportunities for location-enhanced business analytics. The approaches presented in the marketing and operations research literature commonly assume that a business that wants to do analysis owns the data about it. However, this is rarely the case. Location data is typically collected by mobile telecommunication operators and service providers, such as Foursquare. These data owners seek ways to enable other businesses to run location-based analytics queries without violating their customers' privacy. Thus, one needs to prevent the location-based service providers from tracking the users individually, while still allowing other businesses to obtain useful information. Similarly, businesses do not want to share their customer lists with location-based service providers. In this work, we develop efficient privacy-preserving query processing protocols that help to identify the best locations to open new branches considering the distribution of the customer locations.

Optimal location selection is a common location-based analysis that seeks the best location to open a new facility optimizing an objective function given a set of existing facilities and a set of customers. A common approach is to utilize computational geometry techniques on the customer locations with the assumption that the locations are known.

However, third party businesses and analysts cannot use these techniques in real life because customer locations are not always known by these businesses. To perform successful location-based queries, businesses need up-to-date locations that can be gathered from location data owners, such as mobile operators and location-based service providers. For instance, while retail stores or banks may know the home addresses of their customers, they may also like to know their locations during certain time periods in the day. Work addresses of the customers may be missing or out-of-date in their databases. The location information needs to be gathered from data owners while preserving sensitive information of businesses and data owners, as well as the privacy of their customers including their identity and location. 

To be consistent, in this paper we refer the location data owner as the server, and the business that requests queries as the client. We refer their customers as the users of the server and the users of the client. The client has existing facilities, such as branches of a bank, and aims to find the optimal location for the new one among several candidates. The client is able to request a fundamental class of queries that can be used in optimal location selection. In these queries, the client only obtains aggregate information about locations of its users without learning the location of any specific user. The client has several candidates for the new facility and it can request the queries for each candidate location and select the best one. 

A simple example to these aggregate queries is average distance query, in which the client retrieves the average distance of its users to their nearest facilities. The nearest facility of each user is the facility that has the minimum distance to that user. The average distance is a valuable information for the client to minimize it for maximizing user benefit. In a non-privacy-preserving solution for this query, the client sends the facility locations and its user list to the server. The server checks the location of each user (who gave informed and explicit consent for this information) and calculates distances to their nearest facilities. At the end of the query, the server returns the average distance and the client obtains useful information for facility location without tracking its users individually. The client can send a different location for the new facility in each query together with the locations of existing facilities. As a result, it can select the best candidate that minimizes the average distance between users and their nearest facilities.

For a privacy-preserving solution, we need to hide the client's user list and the server's user list from each other. We also need to hide the answer to the query from the server. Otherwise, the server learns the best candidate for the new facility and it may share this information with the competitors. We investigate privacy-preserving solutions to aggregate queries which allow analyzing location data in servers and selecting the best facility location. With the proposed solutions, without sharing its user list with the server, the client can obtain aggregate information about user locations and find an optimal place for its new facility among several candidates depending on different objective functions. These objectives are (i) uniformly distributing the cardinality of the reverse nearest neighbors (RNN), i.e., the set of points that has the query point as the closest facility, (ii) minimizing the average distance between each user and her closest facility, and (iii) minimizing the maximum distance between a user and her closest facility.

We define three fundamental aggregate queries for optimal location selection and propose two types of privacy-preserving query-processing protocols for each type of query, utilizing partially homomorphic encryption as a building block. We encrypt the sensitive data of the server and the client, and perform the operations on the encrypted data to preserve the privacy of both parties. First, we explain server-based protocols, in which most computation is performed by the server, and hence the workload of the client is low. This solution is particularly convenient when the client has limited computational power. To decrease the communication overhead in each query, we also propose client-based protocols. In these protocols, the client performs the majority of the computation during the setup phase (which occurs only once). After completion of the setup phase, all queries are processed with low communication overhead. Therefore, our client-based solution is highly efficient when the client undertakes some pre-computations before running its queries.

During the protocols, homomorphic encryption is used for keeping the user list of the client and the query result hidden from the server and keeping the user list of the server and location data hidden from the client. Initially, we describe the protocols to return exact query results. Since the server is unaware of the query result and the queries return aggregate results, some queries may leak information about users. For instance, if the result of a counting query is one, that user can be predicted by the client. To prevent information leak about any single user, we also satisfy differential privacy in our protocols by adding controlled noise to the query result. Therefore, we use homomorphic encryption and differential privacy together to guarantee privacy of individuals during query processing. Our contributions are summarized as follows:

\begin{enumerate}

	\item We introduce a practical setting in which the client (e.g., a business) runs a useful class of location-based queries on the database of the server (e.g., a location-based service provider) without violating the privacy of individuals involved both in the client and the server side.
	\item We enhance facility location problems by removing the assumption that the customer locations are known to the businesses. With the proposed solutions, a business can find the best location for a new facility among several candidates without knowing its customer locations.
	\item We introduce two novel query processing protocols for different types of queries, i.e.,  RNN cardinality query, average distance query, and maximum distance query that can be used as a service to identify optimal facility location. Our protocols utilize homomorphic encryption for protecting privacy of both parties and satisfy differential privacy. We also discuss the impact of differential privacy on the utility of the protocols.
	\item The proposed protocols take advantage of using a potential superset of user space to hide the user lists of both parties. Our solution does not use any computationally expensive cryptographic comparisons such as private equality testing or private set intersection. The performance evaluations show that the proposed protocols are practical, efficient, and scalable. For instance, when the server has 25 million users, executing privacy-preserving RNN cardinality query takes around 10 seconds on a modest computer.

\end{enumerate}

The remainder of this paper is organized as follows: A literature review and background information are given in Section~\ref{related}. Section~\ref{sec:problem} presents the system model, the threat model, and the definitions of the aggregate queries for optimal location selection. We describe the server-based solutions in Section~\ref{sec:server} and the client-based solutions in Section~\ref{sec:client}. In Section~\ref{difPriv}, we explain how to achieve differential privacy in our protocols. We present our experimental results in Section~\ref{eval}. Finally, we conclude in Section~\ref{conc}.

\vspace{-3mm}
\section{Related Work and Background}
\label{related}

Since our work is related to optimal location queries and privacy-preserving location-based query processing, we give the literature review of both subjects and explain the major differences between our work and previous works in the literature. The concept of differential privacy and homomorphic encryption schemes are also explained in this section as building blocks of our protocols.

\vspace{-3mm}
\subsection{Optimal Location Queries}
\label{sec:opt}

Given a set of existing facilities and a set of users, the optimal location query \cite{du2005optimal} finds a location $l$ for the new facility with maximum influence. The influence of a point is commonly formalized based on its RNNs \cite{korn2000influence}. The RNN query finds the set of points that has the query point as the nearest neighbor (NN). There are two variants of RNN queries. In the monochromatic version, all points belong to the same category. In the bichromatic version, points are divided into two categories, such as users and facilities. Given a facility $f$, the bichromatic RNN query finds the set of users that has $f$ as the nearest facility. The general assumption in optimal location queries is that each user prefers her closest facility. Therefore, the RNN query plays an important role in facility location problems because a facility's RNN is the set of users who prefers this facility.

Businesses run optimal location queries to find the best locations for their new facilities. The definition of \textquotedblleft best location" or \textquotedblleft location with maximum influence" depends on the type of the facility. In \cite{du2005optimal}, the influence of a location is defined as the total weight of its RNNs. The authors define the problem with weighted users and aim to maximize the total weight of users that are closer to the new location than to their closest facilities. $L_1$ distance is considered in \cite{du2005optimal} and they propose three methods to solve the problem. Another solution to maximize the bichromatic RNN for $L_2$ distance is proposed in \cite{wong2009efficient}.

In the literature, there are also other definitions of the \textquotedblleft best location" which aim to maximize user benefit and increase service quality. One of them is minimizing the maximum distance between a user and her closest facility \cite{cardinal2006min, chen2014efficient}. Another objective is minimizing the average distance between each user and her closest facility. The problem is proposed as min-dist optimal-location query in \cite{zhang2006progressive}. This query has many real-life applications where it aims to improve the quality of service or reduce the logistics cost by businesses. \cite{zhang2006progressive} and \cite{qi2014min} solve the problem with $L_1$ and $L_2$ distance assumptions, respectively.

In previous works on facility location problems, it is assumed that customer locations are known. In this paper, we assume that customer locations are not known by businesses, but stored in a location-based service provider, and businesses need to analyze location data by requesting queries. We define three aggregate queries for optimal location selection and develop privacy-preserving protocols for them. These queries are defined to analyze the location data and they can be used in optimal location selection. Businesses can decide the best location among the candidates by requesting several queries and comparing the query results.

\vspace{-3mm}
\subsection{Privacy-Preserving Location-Based Query Processing}
\label{sec:locpriv}

Today, vast amounts of information are collected and analyzed in databases around the world. Data may be stored by multiple parties and these parties may not be keen on sharing their data with others. In secure multi-party computation (SMC), multiple parties jointly compute a function over their inputs without revealing their inputs to each other. In \cite{du2001secure}, several SMC problems are identified. One such problem defined in \cite{du2001secure} is the privacy-preserving database query, where Alice seeks a match with her private string $q$ in Bob's database $T$. The privacy requirement is hiding $q$ and the query result from Bob, and hiding $T$ from Alice. The authors develop an efficient solution for the matching problem in \cite{du2001protocols} by using a semi-trusted third party.

Privacy-preserving location-based queries have been studied in the literature. Cheng et al. \cite{cheng2006preserving} propose a privacy-preserving range query protocol to find users within a range with non-zero probability. In~\cite{cheng2006preserving}, each user has a cloaked region to hide her exact location, and the probability of being within a range depends on the intersection of the cloaked regions. A hybrid approach that integrates private set intersection and location cloaking is presented in \cite{wu2013hybrid}. For privacy-preserving NN queries, a privacy-aware query processing framework called Casper is presented in \cite{mokbel2006new}. This framework uses a location anonymizer to blur users' exact locations into cloaked regions. Ghinita et al. \cite{ghinita2008private} eliminate the usage of third-party anonymizers by using cryptographic techniques. They utilize private information retrieval techniques to preserve location privacy. In \cite{qi2008efficient}, efficient protocols are proposed for privacy-preserving k-NN searches by using several primitive SMC protocols. Yi et al. \cite{yi2014practical} present solutions for the same problem and use Paillier encryption and location cloaking as building blocks. 

For privacy-preserving location-based query processing, one can follow several approaches, such as location perturbation \cite{mokbel2006new}, providing k-anonymity by dummy locations \cite{niu2014achieving}, data transformation \cite{khoshgozaran2007blind}, and using cryptography \cite{ghinita2008private, qi2008efficient, yi2014practical}. We follow the cryptographic approach, which provides privacy without compromising utility. However, providing exact query results may cause information leaks in some cases such as counting queries. Therefore, we integrate the principle of differential privacy \cite{dwork2008differential} into the proposed protocols. We explain the notion of differential privacy in Section \ref{sec:difPriv}.

Existing works on location privacy try to hide the location information that the client (i.e. querying side) has from the server (i.e. location-based service provider). In our scenario, user location information is stored in the server and the server hides this sensitive information from the client. The client wants to analyze its customers' locations in order to find the optimal facility location. One approach to allow analytics on the location data can be publishing anonymized data by the server. However, the client cannot identify its users in anonymized data and anoymized location data can also be vulnerable to de-anonymization attacks \cite{de2013unique}. Therefore, the client should retrieve its users' aggregate information via privacy-preserving queries. Since both parties must hide their user lists from each other, the server and the client must find their common users collaboratively without learning these common users. In this work, we propose novel secure two-party protocols that allow analyzing location data in the server. We develop our protocols using potential superset of user space to hide the user lists of both parties.

\vspace{-3mm}
\subsection{Differential Privacy}
\label{sec:difPriv}

Differential privacy aims to protect the privacy of individuals while releasing aggregate information about the database. It is based on the neighborhood of databases. Two databases $\texttt{D}$ and $\texttt{D'}$ are neighbors if they differ in only one entry. Differential privacy requires that query results for two neighbor databases should be indistinguishable. Let the output of a protocol $P$ on database $\texttt{D}$ be $P(\texttt{D})$. The differential privacy is formally defined as follows:

\begin{def1}
Protocol $P$ satisfies $\epsilon$-differential privacy if for any two neighbor databases $\texttt{D}$ and $\texttt{D'}$, and any subset $S$ of output space of $P$,
\begin{displaymath}
\mathrm{Pr}\left[P(\texttt{D}) \in S\right] \leq \mathrm{Pr}\left[P(\texttt{D'}) \in S\right] \cdot e^{\epsilon}
\end{displaymath}
\end{def1}

A typical way to achieve differential privacy is adding controlled random noise to the query result. For numeric queries, Laplace mechanism can be used to produce the noise drawn from the Laplace distribution. Let $Laplace(\lambda)$ be a sample from Laplace distribution with mean 0 and standard deviation $\lambda$. To obtain $\epsilon$-differential privacy, the noise drawn from the Laplace distribution must be calibrated according to the sensitivity of the protocol \cite{dwork2006calibrating}. The sensitivity of the protocol is the maximum possible change on the output by changing a single record in database. Given a protocol $P$, the sensitivity of the protocol is defined as follows:

\begin{def1}
Let $\mathcal{N}$ be the set of all pairs of neighbor databases.
\begin{displaymath}
\Delta P = \max_{(\texttt{D}, \texttt{D'})\in \mathcal{N}} \left\|P(\texttt{D}) - P(\texttt{D'})\right\|
\end{displaymath}
\end{def1}

Therefore, a protocol $P$ satisfies $\epsilon$-differential privacy for the result
\begin{displaymath}
P(\texttt{D}) + Laplace(\frac{\Delta P}{\epsilon})
\end{displaymath}
In Section \ref{difPriv}, we show the sensitivity of each considered query and how to achieve differential privacy during the protocols.

\vspace{-3mm}
\subsection{Homomorphic Encryption}
\label{sec:homom}

In homomorphic encryption, a specific algebraic operation performed on the plaintext is equivalent to another (possibly different) algebraic operation performed on the ciphertext. Cryptosystems that allow homomorphic computation for a limited number of operations such as addition or multiplication are called partially homomorphic. For instance, given two messages $x$ and $y$, one can compute the encryption of $x+y$ by using the encryptions of $x$ and $y$ in an additive homomorphic encryption scheme. In multiplicative homomorphic schemes, $E(x \cdot y)$\footnote{For the rest of the paper, $E(x)$ denotes the encryption of message $x$} can be computed by using $E(x)$ and $E(y)$. Gentry \cite{gentry2009fully} proposed first fully homomorphic encryption scheme that supports both addition and multiplication. Since partially homomorphic schemes are more efficient and calculating the sum is sufficient for our protocols, we are interested in additive homomorphic cryptosystems \cite{benaloh1994dense, okamoto1998new, paillier1999public}, satisfying $E(x) \cdot E(y) = E(x + y)$. Another homomorphic property of these cryptosystems is that encrypted plaintext $E(x)$ raised to a constant $k$ is equal to encryption of the product of the plaintext $x$ and the constant $k$, i.e. $E(x)^k = E(x \cdot k)$.

We develop our protocols by using the Paillier cryptosystem \cite{paillier1999public}. In Paillier, if the public key (PK) is the modulus $m$ and the base $g$, then the encryption of a message $x$ is $E(x) = g^x \cdot r^m ~(\mathrm{mod}~ m^2)$, for some random $r \in \left\{0,...,m-1\right\}$. Using a random value $r$ in encryption ensures that two messages that are the same will encrypt to the same value with only a negligible likelihood. Hence, Paillier provides \emph{semantic security}. $m$ should be selected as the product of two primes $p$ and $q$. The private keys (SK) of the Paillier cryptosystem are $\lambda = lcm(p - 1, q- 1 )$ and $\mu = (L(g^{\lambda} ~ \mathrm{mod} ~ m^2))^{-1} ~ \mathrm{mod} ~ m$, where $lcm(a,b)$ is the least common multiple of $a$ and $b$, and $L(u) = \frac{u-1}{m}$. The decryption of a ciphertext $c$ can be performed using private keys as follows: $D(c) = (L(c^{\lambda} ~ \mathrm{mod} ~ m^2) \cdot \mu) ~ \mathrm{mod} ~ m$. Paillier satisfies $E(x) \cdot E(y) = E(x + y)$, because $(g^x \cdot r_1^m) \cdot (g^y \cdot r_2^m) = g^{x+y} \cdot (r_1 + r_2)^m$. As a result of this homomorphic property, multiplying a ciphertext $E(x)$ with $E(0)$ creates another ciphertext which is the fresh encryption of $x$.

\vspace{-3mm}
\section{Problem Formulation}
\label{sec:problem}

We present our system model in Section \ref{sec:system}. Formal definitions of the queries are given in Section \ref{sec:querydef}. We describe the threat model in Section \ref{sec:threat}.

\vspace{-3mm}
\subsection{System Model}
\label{sec:system}

There is a server ($\mathcal{S}$) (e.g., a location-based service provider) that provides analytics as a service and a client ($\mathcal{C}$) that requests queries. The server is the database owner and has $n_s$ users $\mathcal{U_S} = \left\{S_1, S_2,...,S_{n_s}\right\}$. In addition, the server has location information for each $S_i$ at different time periods. The client has $n_c$ users $\mathcal{U_C} = \left\{C_1, C_2,..., C_{n_c}\right\}$ and a list of its $k$ existing facilities $\mathcal{F} = \left\{F_1, F_2,..., F_k\right\}$. The locations of the existing facilities are public and known by the server. The client wants to run aggregate queries such as count, sum, and maximum on the location data of the server, e.g., to analyze the candidate locations for a new branch. The client aims to hide $\mathcal{U_C}$ and the query results from the server. The server also aims to hide $\mathcal{U_S}$ from the client and prevent user tracking by the client. Hence, the client will not learn anything about the location of any specific user; it will only obtain the query result at the end of the protocol.

\begin{figure}%
\includegraphics[width=\columnwidth]{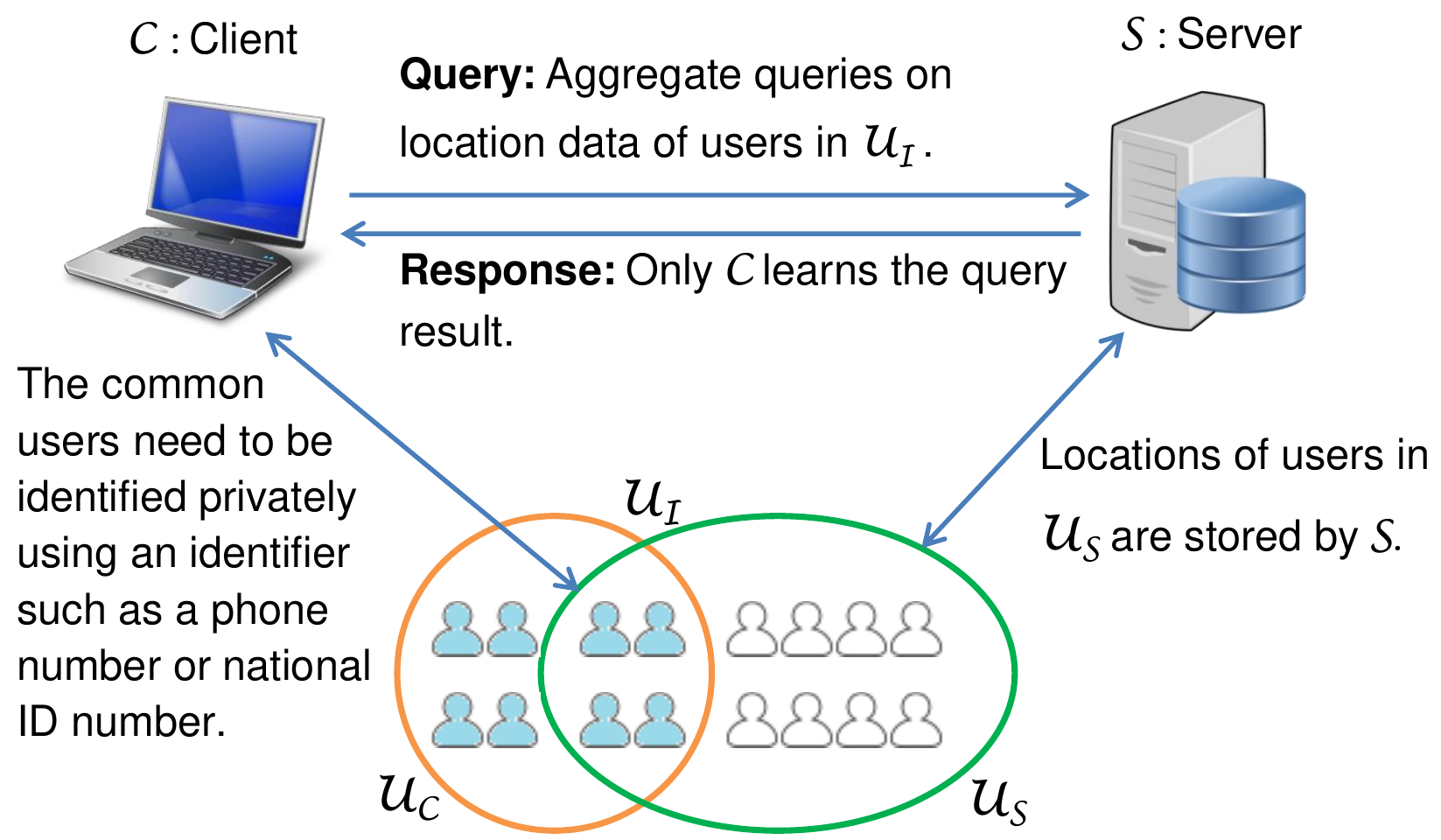}%
\caption{System model.}%
\vspace{-3mm}
\label{sysMod}%
\end{figure}

We sketch out our system model in Figure \ref{sysMod}. To run aggregate queries about its users, the client must identify its users in $\mathcal{U_S}$ using an identifier. Before running queries, the server and the client decide on an identifier such as mobile phone number. Most businesses and service providers know mobile phone numbers of their customers. Another identifier can be national identification number. If the server is a telecommunication company and the client is a bank or a hospital they might use national identification number as the identifier. Let $\mathcal{U_I}$ be $\mathcal{U_S} \cap \mathcal{U_C}$, and $n_I$ be the cardinality of $\mathcal{U_I}$. Since the server does not have the location information of users in $\mathcal{U_C} \backslash \mathcal{U_S}$, we define our queries for the users in $\mathcal{U_I}$.

We define three useful types of queries for this context: RNN Cardinality Query (RNNQ), Average Distance Query (AVGQ), and Maximum Distance Query (MAXQ). Since the server knows the user locations, it can calculate the distance between a user and a facility via any distance measure. The main challenges are keeping $\mathcal{U_C}$ hidden from the server and preventing user tracking by the client.

We propose two types of solutions for each query type, the server-based solutions and the client-based solutions. The server is responsible for most of the computation in the server-based solutions. Hence, they are suitable when the client prefers outsourcing computation. The drawback of server-based solutions over the client-based version is their communication overhead. The client-based solutions reduce communication overhead significantly. In the client-based solutions, most of the computation is performed by the client only in the setup phase. In Section \ref{sec:server} and Section \ref{sec:client}, we describe the server-based and the client-based protocols which return exact query results. Since exact query results may leak information in some cases such as counting queries, in Section \ref{difPriv} we explain how to add controlled random noise to the query result in each protocol to satisfy differential privacy.

\vspace{-3mm}
\subsection{Query Definitions}
\label{sec:querydef}

\subsubsection{RNN Cardinality Query (RNNQ)}
\label{sec:rnn}

One of the objectives of optimal location queries is uniformly distributing the workload in facilities. In this case, the new facility should attract users from dense facilities. Attracting a user is equivalent to being the closest facility to the user. This query finds the number of users attracted by each facility. The formal definition of the RNNQ is as follows:

\begin{qry}
Given facility locations, find the total number of users in $\mathcal{U_I}$ attracted by each facility. In other words, calculate the cardinality of RNN for each facility.
\end{qry}

In practice, the client can initially run the RNNQ with existing facilities $\mathcal{F}$ to analyze the distribution of the users. Using the result, the client can determine candidate locations for the new facility $F_{k+1}$. For candidate locations, the client can run the RNNQ with $\mathcal{F} \cup F_{k+1}$. Hence, the client can observe the total number of users attracted by each candidate location for $F_{k+1}$ and select the location that provides the most balanced distribution.

\vspace{-2mm}
\subsubsection{Average Distance Query (AVGQ)}
\label{sec:avgdist}

One of the objectives of optimal location queries is minimizing the average distance between each user and her closest facility. For instance, delivery services pay attention to decreasing the average distance between their customers and the nearest shop. The AVGQ is formalized as follows:

\begin{qry}
Given facility locations, find the average distance between users in $\mathcal{U_I}$ and each one's nearest facility.
\end{qry}

In practice, the client can run the AVGQ with $\mathcal{F} \cup F_{k+1}$, where $F_{k+1}$ is a candidate location for the new facility. Hence, the client can select the optimal location for $F_{k+1}$, which minimizes the average distance.

\vspace{-2mm}
\subsubsection{Maximum Distance Query (MAXQ)}
\label{sec:maxdist}

Another objective of optimal location queries is minimizing the maximum distance between a user and her closest facility. In this objective, the aim is to optimize the worst-case cost of reaching the nearest facility. The MAXQ is formalized as follows:

\begin{qry}
Given facility locations, find the maximum distance between a user in $\mathcal{U_I}$ and her nearest facility.
\end{qry}

In practice, the client can run MAXQ with $\mathcal{F} \cup F_{k+1}$, for candidate $F_{k+1}$ locations. The client can select the optimal location for $F_{k+1}$, which minimizes the maximum distance.

\vspace{-3mm}
\subsection{Threat Model}
\label{sec:threat}

In our model, both the server and the client are considered \textquotedblleft semi-honest\textquotedblright. Therefore, both parties follow the protocol correctly; however, they may try to learn additional information by analyzing the data. That is, the server may try to determine the client's user list, and similarly, the client may try to determine the individual locations of its users during the protocol (by using the messages they receive throughout the protocol). On the other hand, both the server and the client follow protocol execution honestly by forming correct messages, input, and output parameters for each other. This is a reasonable assumption in the problem setting since both parties are motivated to produce the correct result. The server sells the service and the correct result increases the client's satisfaction. Also, the client finds the best facility location if the query results are correctly calculated.

The proposed solutions are secure two-party protocols in which the server and the client wish to compute the query result securely without sharing their inputs with the opposing party. Both the server and the client have sensitive data that should be hidden from the other party. We formally list the sensitive data as follows:
\begin{enumerate}
	\item Input of the client: $\mathbfcal{U_C}$.
	\item Input of the server: (a) $\mathbfcal{U_S}$ and (b) \textbf{location information of users in } $\mathbfcal{U_S}$. 
	\item Output of the protocol: \textbf{Query result.} 
\end{enumerate}

We aim to hide all of the above sensitive data (from unauthorized parties) in our protocols. The parties must not learn the input of each other. At the end of the protocols, only the client must get the query result and the server must not learn it. The privacy of the server is assured if sensitive data 2 is hidden from the client, and the privacy of the client is assured if sensitive data 1 \& 3 are hidden from the server. We prove the security of our proposed protocols in the semi-honest model using the simulation paradigm defined in \cite{goldreich2009foundations}.  

While the locations of existing facilities are typically public, the location of a new facility can be sensitive data for the client. In this case, the client can run the query with some dummy locations to provide $K$-anonymity \cite{niu2014achieving}, which provides indistinguishability among $K$ locations. Since the query result is hidden from the server, all of the $K$ locations are indistinguishable for the server.

One potential threat to the server's sensitive data may be obtaining information via exhaustive client queries. By using non-existing facilities, the client can try to obtain information about location of some users. For instance, the client can divide the whole region into two regions and select the center of each region as a facility location. When the client performs RNNQ with these facility locations, it learns the total number of users in each region. The client can divide each region into smaller regions in subsequent queries, until each region has at most one user. At the end, the client learns the small regions which contains a user and it may predict the user in a small region with background knowledge. Therefore, if the total number of facilities in the query is very small or very large, the client may obtain information about user locations.

We assume the locations of $k$ existing facilities of the client are public and known by the server. The server decides two threshold values $\theta_1$ and $\theta_2$ such that the client can add at most $\theta_1$ new facilities or remove at most $\theta_2$ existing facilities in a query\footnote{$\theta_1$ and $\theta_2$ are design parameters of RNNQ, AVGQ, and MAXQ to be decided by the server.}. Thus, when the client sends the locations of the facilities, the server aborts the protocols in following cases:
\begin{itemize}
	\item if the total number of facilities is greater than $k + \theta_1$,
	\item if the total number of facilities is less than $k - \theta_2$,
	\item if the facilities in the query do not include at least $k - \theta_2$ existing facilities of the client.
\end{itemize}

There is a tradeoff between utility and privacy in the selection of these threshold values. Selecting small $\theta_1$ and $\theta_2$ increases privacy, however, the utility of the protocols decreases due to rejection of more queries. Therefore, there cannot be an optimal threshold value for the protocols.

Moreover, when the query result includes a small number of users, the client can make an estimate about these users. For instance, there may be only one user whose nearest facility is a particular facility in RNNQ. Hence, if the RNN cardinality of a facility is one in RNNQ, the client can predict that user using its background knowledge. To prevent such privacy leaks in our protocols, we explain how to provide differential privacy in Section \ref{difPriv}.

Finally, we also assume that during the protocol, communication is encrypted between the server and the client against an eavesdropper and that the server and the client(s) do not collude.

\vspace{-3mm}
\section{Server-based Query Processing Protocols}
\label{sec:server}

In this section, we propose server-based solutions that preserve the privacy while processing the queries in Section \ref{sec:querydef}. We introduce the high-level overview of the server-based protocols in this section and we give the detailed steps of the protocols in Appendix. We present the security analysis of server-based protocols in Section \ref{sec:evalServ}. Table~1 shows the symbols used in the protocols.

\begin{table}
  \centering
  \caption{Symbols used in protocols.}
	\vspace{-2mm}
  \begin{tabular}{|c|c|} \hline
   $m_s$, $m_c$ & modulus in Paillier generated by ($\mathcal{S}$, $\mathcal{C}$)\\ \hline
    $g_s$, $g_c$ & base in Paillier generated by ($\mathcal{S}$, $\mathcal{C}$)\\ \hline
		$PK_s$, $PK_c$ & public keys of $\mathcal{S}$ and $\mathcal{C}$\\ \hline
		$SK_s$, $SK_c$ & private keys of $\mathcal{S}$ and $\mathcal{C}$\\ \hline
		$E_s(x)$, $E_c(x)$ & Encryption of message $x$ using ($PK_s$, $PK_c$)\\ \hline
		$\left[x\right]_s$, $\left[x\right]_c$ & denotes $x$ is encrypted using ($PK_s$, $PK_c$)\\ \hline
		$D_s(\left[x\right]_s)$, $D_c(\left[x\right]_c)$ & Decryption of ciphertext $x$ using ($SK_s$, $SK_c$)\\ \hline	
		$d(a,b)$ & Distance between points $a$ and $b$\\ \hline
    $\mathcal{U_S}$, $\mathcal{U_C}$ & user sets of $\mathcal{S}$ and $\mathcal{C}$\\ \hline
    $\mathcal{U}$ & superset of $\mathcal{U_S}$ and $\mathcal{U_C}$\\ \hline
    $\mathcal{U_I}$ & $\mathcal{U_S} \cap \mathcal{U_C}$\\ \hline
    $n$, $n_s$, $n_c$, $n_I$ & total number of users in ($\mathcal{U}$, $\mathcal{U_S}$, $\mathcal{U_C}$, $\mathcal{U_I}$)\\ \hline
    $\mathcal{F}$ & set of existing facilities of $\mathcal{C}$\\ \hline
    $k$ & total number of existing facilities\\\hline
    $q$, $\mathcal{Q}$ & result (value, set) of the query\\\hline
		$w$ & random number greater than $q$ in MAXQ\\
    \hline\end{tabular}
		\vspace{-4mm}
\end{table}

The underlying protocols utilize the additive homomorphic property to hide sensitive data from other parties by calculating the sum of the encrypted values without decrypting them. We utilize Paillier cryptosystem as an additive homomorphic scheme satisfying $E(x) \cdot E(y) = E(x + y)$. In the server-based protocols, the server creates a public and private key pair ($PK_s$, $SK_s$), and shares the public key with the client. The client can encrypt any value or perform homomorphic operations on the ciphertexts, but only the server can decrypt encrypted messages. The server performs the majority of the encryptions in the protocols.

In the setup phase, the server generates ($PK_s$, $SK_s$) for Paillier cryptosystem. In addition, the server selects a superset $\mathcal{U} = \left\{U_1,...,U_n\right\}$ of $\mathcal{U_S}$ such that $\mathcal{U_S} \subset \mathcal{U}$. The aim of selecting $\mathcal{U}$ is hiding $\mathcal{U_S}$ (sensitive data 2(a)) from the client. For instance, let the identifier used in the protocols be mobile phone numbers. Location-based service providers such as Foursquare and mobile telecommunication operators, and most businesses such as banks, hotels, and retailers typically know the mobile phone numbers of their customers. Hence, they can use mobile phone numbers as identifiers. Assume the phone numbers consist of 7 digits and there are 50 different mobile operator codes. When the superset $\mathcal{U}$ contains all possible mobile phone numbers, $n$ becomes 500 million. Since $\mathcal{U}$ contains all possible numbers, it completely protects $\mathcal{U_S}$ from the client. Another example is using national identification numbers as identifier. If national id numbers consist of 9 digits and the superset $\mathcal{U}$ contains all possible id numbers, $n$ becomes one billion. The server shares $PK_s = (g_s, m_s)$ and $\mathcal{U}$ with the client. Note that all multiplications and exponentiations of ciphertexts in the server-based protocols are calculated in $\mathrm{mod}~m_s^2$.

\begin{figure}%
\includegraphics[width=\linewidth]{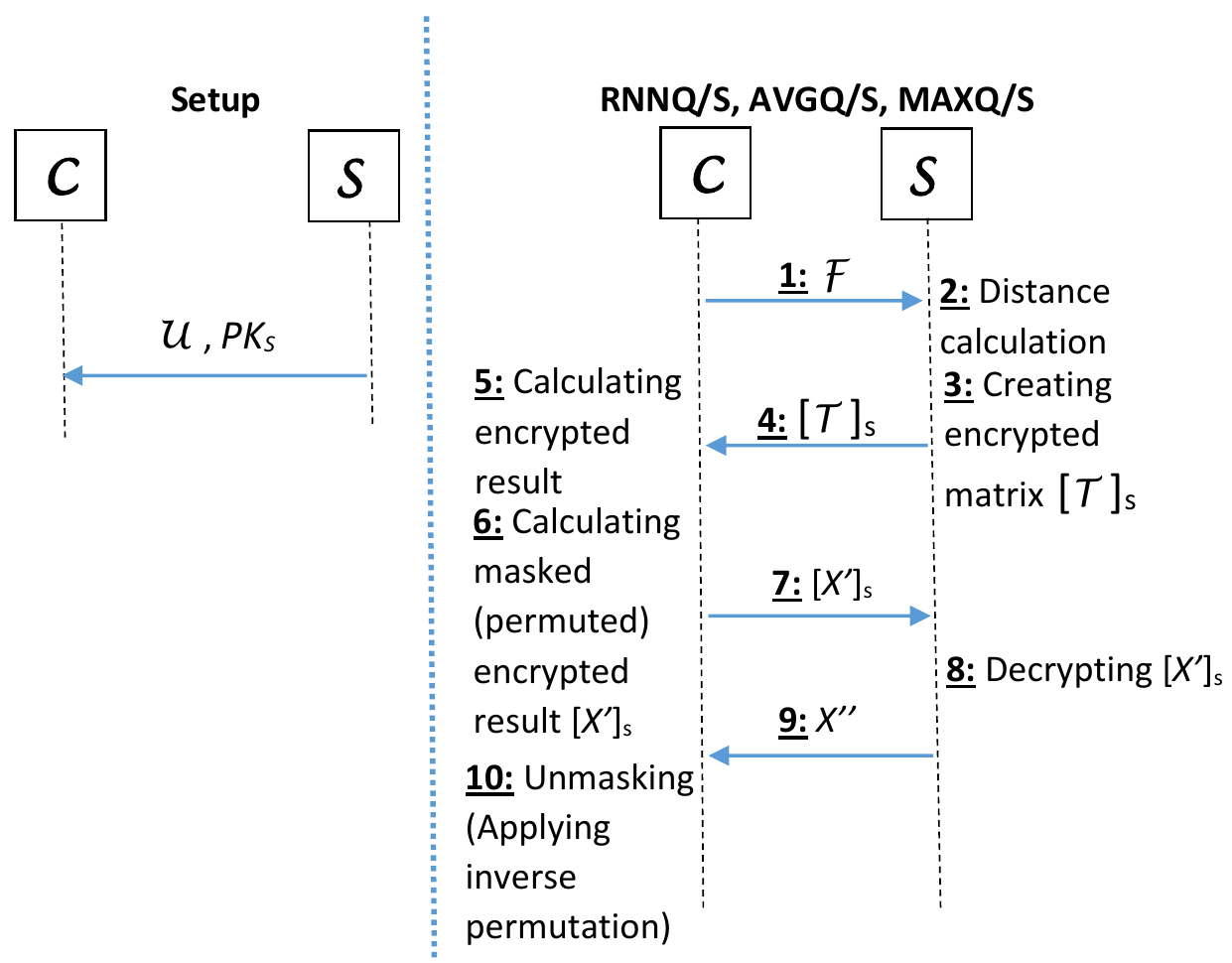}%
\caption{Overview of the server-based protocols.}%
\vspace{-3mm}
\label{prt1}%
\end{figure}

Figure~\ref{prt1} shows the overview of the setup phase and the protocols. Here, we briefly explain the steps of the server-based solutions and illustrate these steps with an example scenario for RNNQ/S. The server-based protocols consist of 10 steps. Steps 1, 4, 7, and 9 are the communication steps. In the first step, the client sends the query and the facility locations ($\mathcal{F}$) to the server. Step 2 is the calculation of distances between facilities and users. The server determines the nearest facility for each user. Since encrypted values cannot be decrypted by the client, the server computes encrypted values based on nearest facility of each user in Step 3 to hide $\mathcal{U_S}$ and user locations (sensitive data 2(a) \& 2(b)) from the client. Using the encrypted values, the client calculates the ciphertext of the query result by utilizing homomorphic properties of Paillier cryptosystem in Step 5. To hide $\mathcal{U_C}$ and the query result (sensitive data 1 \& 3) from the server, the client masks the encrypted query result in Step 6 before sending to the server for decryption. The server decrypts the encrypted masked result in Step 8 and obtains the masked result. Due to masking in Step 6, the server cannot deduce the query result. In Step 10, the client applies unmasking and finds the query result.

Let the identifier used by the server and the client consists of one digit, and id numbers of the users of the server be $1,3,5,6,7,9$. The server can select the superset $\mathcal{U} = \left\{0, 1, 2, 3, 4, 5, 6, 7, 8, 9\right\}$ such that $\mathcal{U_S} \subset \mathcal{U}$. Assume that we have two facilities $F_1$ and $F_2$. When the client requests RNNQ/S, the server determines the nearest facility of its 6 users. Let $F_1$ be the nearest facility of the users $1$, $6$, and $9$, $F_2$ be the nearest facility of the users $3$, $5$, and $7$. In Step 3, the server computes $\left[T_1\right]_s = \left\{ E_s(0),~E_s(1),~E_s(0),~E_s(0),~E_s(0),~E_s(0),~E_s(1),~E_s(0), \right. $ $\left. E_s(0),~E_s(1)\right\}$ for $F_1$ and $\left[T_2\right]_s = \left\{E_s(0),~E_s(0),~E_s(0), \right. $  $\left. E_s(1),~E_s(0),~E_s(1),~E_s(0),~E_s(1),~E_s(0),~E_s(0)\right\}$ for $F_2$. The server sends these encrypted values $\left[\mathcal{T}\right]_s$ to the client in Step 4. Let id numbers of the users of the client be $1,2,3,5$. In Step 5, the client calculates two ciphertexts for two facilities by multiplying the ciphertexts of its users in $\left[\mathcal{T}\right]_s$. That is, $\left[x_1\right]_s = \left[T_{1,1}\right]_s \cdot \left[T_{1,2}\right]_s \cdot \left[T_{1,3}\right]_s \cdot \left[T_{1,5}\right]_s$ and $\left[x_2\right]_s = \left[T_{2,1}\right]_s \cdot \left[T_{2,2}\right]_s \cdot \left[T_{2,3}\right]_s \cdot \left[T_{2,5}\right]_s$. These values are the encryption of the query results such as $\left[x_1\right]_s = E_s(1)$ and $\left[x_2\right]_s = E_s(2)$. Let two random values selected by the client in Step 6 be $15$ and $11$. The client encrypts these random values and sends $\left[x'_1\right]_s = \left[x_1\right]_s \cdot E_s(15)$ and $\left[x'_2\right]_s = \left[x_2\right]_s \cdot E_s(11)$ to the server. The server decrypts these values in Step 8 and obtains $x''_1 = 16$ and $x''_2 = 13$. When the client receives these masked values, it subtracts the random values and obtains $q_1 = 1$ and $q_2 = 2$. Therefore, the client learns the RNN cardinality of $F_1$ and $F_2$.

\vspace{-3mm}
\subsection{Security Analysis of Server-Based Protocols}
\label{sec:evalServ}

In this section, we prove the security of the server-based protocols in the semi-honest model. Semi-honest parties follow the protocol correctly; however, they may try to learn additional information by analyzing the messages they receive throughout the protocol. In general, in secure two-party protocol, the goal of the parties is to compute a desired output pair $f(x,y) = (f_1(x,y),f_2(x,y))$ from their inputs $x$ and $y$ without revealing them to each other. The first party wants to obtain $f_1(x,y)$ and the second party wants to obtain $f_2(x,y)$ at the end of the protocol. During the protocol, the view of a party consists of its input, its random-tape, and sequence of incoming messages throughout the protocol. A protocol privately computes $f(x,y)$ if a party's view can be simulated from its input and output\cite{goldreich2009foundations}.

More formally, let $\Pi$ be a secure two-party protocol for computing $f(x,y)$. The views of the parties are denoted as \textsc{VIEW}$^\Pi_1(x,y)$ and \textsc{VIEW}$^\Pi_2(x,y)$. Then, the security of a deterministic protocol in semi-honest model is defined as follows \cite{goldreich2009foundations}: 
\begin{def1}
The protocol $\Pi$ privately computes $f(x,y)$ if there exist probabilistic polynomial-time simulators $Sim_1$ and $Sim_2$ such that
\begin{displaymath}
	\left\{Sim_1(x,f_1(x,y))\right\} \stackrel{c}{\equiv} \left\{\textsc{VIEW}^\Pi_1(x,y)\right\}
\end{displaymath}
\begin{displaymath}
	\left\{Sim_2(x,f_2(x,y))\right\} \stackrel{c}{\equiv} \left\{\textsc{VIEW}^\Pi_2(x,y)\right\}
	\end{displaymath}
\end{def1}
where $\stackrel{c}{\equiv}$ implies computational indistinguishability. Therefore, a party's privacy is guaranteed if there exists a simulator that can generate a view indistinguishable from the view of the opposing party. In the following, we prove the security of the server-based protocols using this simulation paradigm.

Let the client be the first party and the server be the second party in our protocols. The private input $x$ of the client is $\mathcal{U_C}$ and the private input $y$ of the server is $\mathcal{U_S}$ and the user locations. $\mathcal{F}$ is also the input of the protocol, which is commonly known by the server and the client. As discussed in Section \ref{sec:threat}, it should not be hidden from the server to prevent attacks via exhaustive client queries. In addition, $PK_s$, $PK_c$, and $\mathcal{U}$ are also known by the server and the client as background information. As discussed before, $\mathcal{U}$ is the superset of the users for keeping the user list of parties from each other. The client should get \textbf{query result} as $f_1(x,y)$ at the end of the protocol while the server receives no output (i.e. $f_2(x,y) = \bot$). 

Since the steps of the server-based protocols are similar as shown in Figure~\ref{prt1}, we consider RNNQ/S in the proof. The security of the other protocols can be proved similarly. In RNNQ/S, $\mathcal{Q} = \left\{q_1,...,q_k\right\}$ is the query result where $q_i$ is the total number of users in $\mathcal{U_I}$ whose nearest facility is $F_i$. Therefore, the view of the client ($\textsc{VIEW}_1$) consists of $\mathcal{U_C}$, $\mathcal{F}$, $\left[\mathcal{T}\right]_s$, and $\mathcal{Q}$. To prove that the server's privacy is assured in the protocol, we need to show that there exists a probabilistic polynomial-time simulator $Sim_1$ such that $Sim_1(\mathcal{U_C}, \mathcal{F}, \mathcal{Q})$ is computationally indistinguishable from $\textsc{VIEW}_1$. Since $\left[\mathcal{T}\right]_s$ contains $n \cdot k$ Paillier ciphertexts, $Sim_1$ can generate $n \cdot k$ random numbers between $0$ and $m_s^2$ and these numbers are computationally indistinguishable from the ciphertexts in $\left[\mathcal{T}\right]_s$ due to the semantic security of Paillier cryptosystem.

On the other hand, the view of the server ($\textsc{VIEW}_2$) consists of $\mathcal{U_S}$, user locations, $\mathcal{F}$, and $X''$. To prove that the client's privacy is assured in the protocol, we need to show that there exists a probabilistic polynomial-time simulator $Sim_2$ such that $Sim_2(\mathcal{U_S},$ user locations$,~\mathcal{F})$ is computationally indistinguishable from $\textsc{VIEW}_2$. This is satisfied by letting $Sim_2$ generate $k$ random numbers between $0$ and $m_s$ to simulate $X''$ because $X''$ contains $k$ values $\left\{q_1 + v_1,...,q_k + v_k\right\}$ where each $v_i$ is a randomly selected number by the client. Thus, we conclude that RNNQ/S protocol securely processes RNN Cardinality queries in semi-honest model.

Although the server-based protocols preserve privacy in semi-honest model, they can be vulnerable to the attack of a malicious client. A malicious client can calculate the encrypted result in Step 5 for a specific customer $U_i$. Therefore, the client can obtain information about the location of $U_i$ such as the nearest facility of $U_i$ and its distance to the nearest facility. However, in any case, it is not possible to find the exact location of $U_i$. To prevent the defined attack by malicious clients while providing the exact query result, we propose client-based protocols in Section \ref{sec:client}. Moreover, Section \ref{difPriv} explains satisfying differential privacy in server-based protocols. To protect the privacy of individuals from these kinds of attacks, differential privacy gives a guarantee that presence or absence of an individual will not affect the final output of the algorithm significantly. When the queries return noisy results instead of exact results, a malicious client cannot obtain the nearest facility of a specific user $U_i$ and its distance to the nearest facility. For instance, let $F_1$ be the nearest facility of $U_i$. Then, the exact query result is $(1,0,0,...,0)$ for the defined attack. However, adding a noise to each of these values will prevent the information leak about $U_i$. Therefore, differential privacy provides privacy guarantees against such attacks from the malicious client.

\vspace{-3mm}
\section{Client-based Query Processing Protocols}
\label{sec:client}

In the protocols defined in Section \ref{sec:server}, the data is encrypted with the public key of the server. The server computes most of the encryptions, which dominates the computation cost. In this section, we propose protocols using the public and private keys ($PK_c$, $SK_c$) of the client, where the client computes the majority of the encryptions, however, instead of performing encryptions during each query, the client performs encryptions in the setup. This makes the setup phase of these protocols more costly than the protocols in Section \ref{sec:server}, however, query processing in these protocols is more efficient in terms of computation and communication costs. The protocols defined in this section also return exact query results as in Section \ref{sec:server}. We describe achieving differential privacy during the client-based protocols in Section \ref{difPriv}.

In the setup phase, the client generates a public and private key pair ($PK_c$, $SK_c$) for Paillier cryptosystem. The client shares $PK_c = (g_c,m_c)$ with the server. All multiplications and exponentiations of ciphertexts in the client-based protocols are calculated in $\mathrm{mod}~m_c^2$. In addition, the server selects a superset $\mathcal{U} = \left\{U_1,...,U_n\right\}$ and shares with the client, as described in Section \ref{sec:server}. Then, for each $U_i \in \mathcal{U}$, the client calculates $\left[T_i\right]_c = E_c(0)$ if $U_i \notin \mathcal{U_C}$ and $\left[T_i\right]_c = E_c(1)$ if $U_i \in \mathcal{U_C}$. The client sends $\left[\mathcal{T}\right]_c = \left\{\left[T_1\right]_c,...,\left[T_n\right]_c\right\}$ to the server. Let $r_i$ be the random number used in the calculation of $\left[T_i\right]_c$. To prevent malicious client attack described in Section \ref{sec:evalServ}, the client sends the total number of its users ($n_c$) and $r = \prod_{i=1}^{n} r_i$ to the server. The server multiplies all $\left[T_i\right]_c$ values and obtains a ciphertext which should be equal to encryption of $n_c$. That is, $E_c(n_c) = g_c^{n_c} \cdot r^m = \prod_{i=1}^{n} \left[T_i\right]_c ~(\mathrm{mod}~ m_c^2)$. The server encrypts $n_c$ with the random value $r$ and verifies the total number of the client's users. If $\prod_{i=1}^{n} \left[T_i\right]_c$ is not equal to $E_c(n_c)$ or $n_c$ is less than a threshold value, the server aborts the protocol. Therefore, a malicious client cannot get the query result for a specific user.

Once the client sends $n$ ciphertexts to the server, any of the aforementioned queries can be performed with small computation and communication overheads. We can assume that the users of the client do not change frequently. Small number of changes on the user list do not have a notable effect on query results as well. Hence, the client can update the encrypted list $\left[\mathcal{T}\right]_c$, when there is a significant change on its user list. In addition, when the client decides an update in $\left[\mathcal{T}\right]_c$, it is not necessary to update all values in $\left[\mathcal{T}\right]_c$. The client can only update a subset of users that contains the users to be changed. For instance, if the superset $\mathcal{U}$ includes 100 million users, to change 100 users in $\left[\mathcal{T}\right]_c$, the client can update a subset of $\left[\mathcal{T}\right]_c$ containing one million users instead of all users in $\left[\mathcal{T}\right]_c$.

\begin{figure}%
\includegraphics[width=\linewidth]{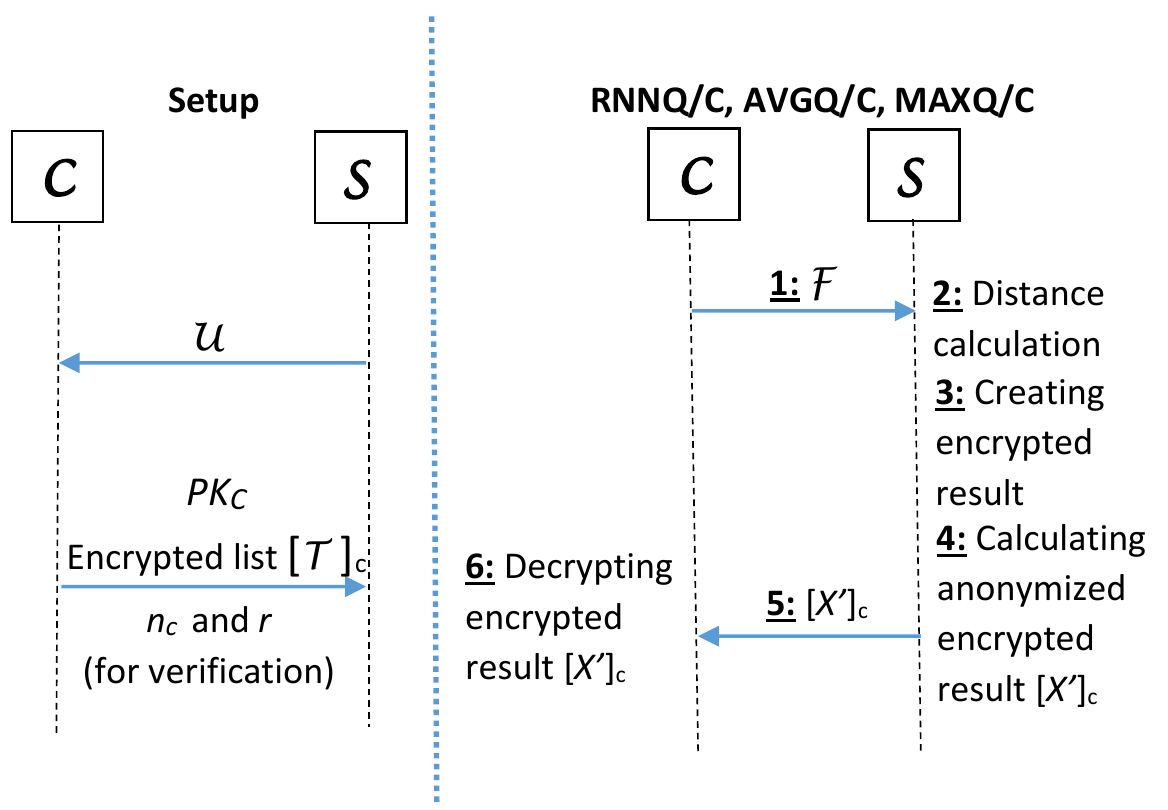}%
\caption{Overview of the client-based protocols.}%
\vspace{-4mm}
\label{prt2}%
\end{figure}

Figure \ref{prt2} shows the overview of the setup phase and the protocols. The protocols in this section consist of 6 steps. The server and the client communicate in Steps 1 and 5. Step 2 is the calculation of distances as in server-based protocols. In Step 3, the server utilizes homomorphic properties of Paillier cryptosystem to calculate the encryption of the query result by using encrypted values in $\left[\mathcal{T}\right]_c$. Before sending the encrypted result to the client, the server anonymizes the result by multiplying it with the encryption of zero in Step 4. This multiplication does not alter the result; it only prevents the server from tracking users by the client. Therefore, the server hides user locations from the client. In Step 6, the client obtains the query result after decryption. Since the server only receives the locations of the facilities during query processing, it is not possible for the server to determine query result. 

\vspace{-3mm}
\subsection{RNN Cardinality Query (RNNQ/C)}
\label{sec:rnn3}

\begin{figure}%
\includegraphics[width=\linewidth]{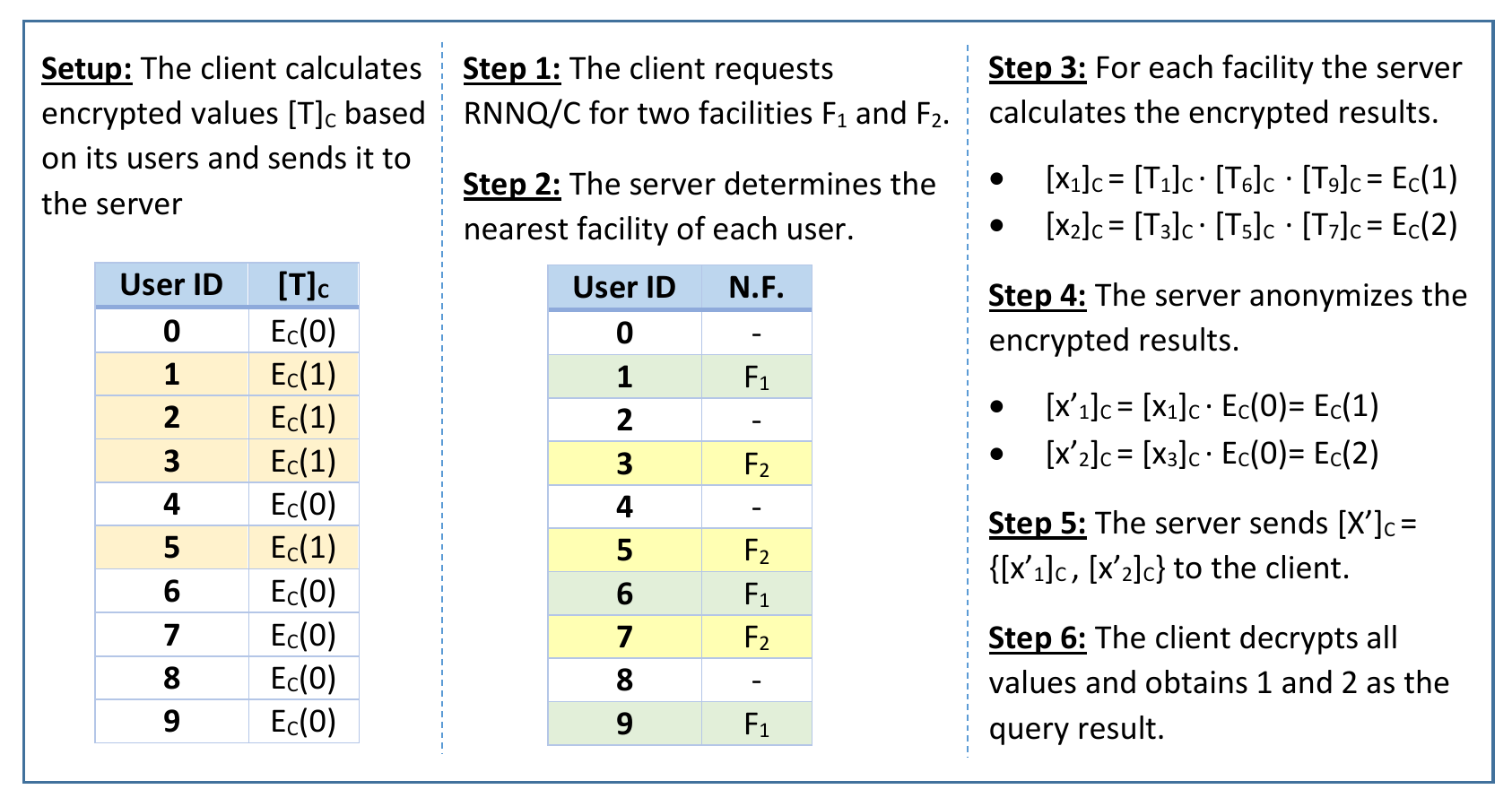}%
\caption{An example scenario for RNNQ/C.}%
\vspace{-3mm}
\label{ex}%
\end{figure}

Let $q_i$ be the total number of users in $\mathcal{U_I}$ whose nearest facility is $F_i$. This query returns the $q_i$ values for each facility $F_i \in \mathcal{F}$. Hence, $\mathcal{Q} = \left\{q_1,...,q_k\right\}$ is the query result. Figure \ref{ex} illustrates the steps of the protocol for the same example scenario explained in Section \ref{sec:server}. The protocol is defined as follows:

\begin{itemize}
\item{\textbf{Step 1:}} $\mathcal{C}$ sends the location of each facility to $\mathcal{S}$.
\item{\textbf{Step 2:}} $\mathcal{S}$ checks the facility locations and aborts the protocol if $\mathcal{C}$ adds more than $\theta_1$ new facilities or removes more than $\theta_2$ existing facilities as described in Section \ref{sec:threat}. $\mathcal{S}$ calculates the distance between each facility and each user in $\mathcal{U_S}$. $\mathcal{S}$ determines the nearest facility of each user $U_i$ in $\mathcal{U_S}$.
\item{\textbf{Step 3:}} For each facility $F_j$, $\mathcal{S}$ calculates the $\left[x_j\right]_c$ value by multiplying $\left[T_i\right]_c$ values such that $U_i \in \mathcal{U_S}$ and the nearest facility of $U_i$ is $F_j$. At the end of this step, $\mathcal{S}$ forms $\left[X\right]_c = \left\{\left[x_1\right]_c,...,\left[x_k\right]_c\right\}$ where $\left[x_i\right]_c$ is the encryption of $q_i$. In this step, $\mathcal{S}$ computes the encrypted result.
\item{\textbf{Step 4:}} $\mathcal{S}$ encrypts 0 using $k$ different random values and calculates $\left[x'_i\right]_c = \left[x_i\right]_c \cdot E_c(0)$ for each $i \in \left\{1,...,k\right\}$.
\item{\textbf{Step 5:}} $\mathcal{S}$ sends $\left[X'\right]_c = \left\{\left[x'_1\right]_c,...,\left[x'_k\right]_c\right\}$ to $\mathcal{C}$.
\item{\textbf{Step 6:}} $\mathcal{C}$ decrypts all $\left[x'_i\right]_c$ values in $\left[X'\right]_c$, and clearly, $D_c(\left[x'_i\right]_c)$ is equal to $q_i$. $\mathcal{C}$ obtains $\mathcal{Q} = \left\{q_1,...,q_k\right\}$.
\end{itemize}

\vspace{-3mm}
\subsection{Average Distance Query (AVGQ/C)}
\label{sec:avgdist3}

Let $q$ be the average distance between users in $\mathcal{U_I}$ and each one's nearest facility. The protocol is defined as follows:

\begin{itemize}
\item{\textbf{Step 1:}} $\mathcal{C}$ sends the location of each facility to $\mathcal{S}$.
\item{\textbf{Step 2:}} As described in RNNQ/C protocol, the server aborts the protocol if it detects a threat. $\mathcal{S}$ calculates the distance between each facility and each user in $\mathcal{U_S}$. $\mathcal{S}$ determines the nearest facility of each user $U_i$ in $\mathcal{U_S}$ and the distance $d_i$ to the nearest facility.
\item{\textbf{Step 3:}} $\mathcal{S}$ calculates the multiplication of $\left[T_i\right]_c^{d_i}$ values and the multiplication of $\left[T_i\right]_c$ values such that $U_i \in \mathcal{U_S}$. That is, $\left[x_1\right]_c = \prod_{U_i \in \mathcal{U_S}} \left[T_i\right]_c^{d_i}$ and $\left[x_2\right]_c = \prod_{U_i \in \mathcal{U_S}} \left[T_i\right]_c$. Clearly, $\left[x_1\right]_c$ is equal to $E_c(q \cdot n_I)$ and $\left[x_2\right]_c$ is equal to $E_c(n_I)$.
\item{\textbf{Step 4:}} $\mathcal{S}$ calculates $\left[x'_1\right]_c = \left[x_1\right]_c \cdot E_c(0)$ and $\left[x'_2\right]_c = \left[x_2\right]_c \cdot E_c(0)$.
\item{\textbf{Step 5:}} $\mathcal{S}$ sends $\left[X'\right]_c = \left\{\left[x'_1\right]_c, \left[x'_2\right]_c\right\}$ to $\mathcal{C}$.
\item{\textbf{Step 6:}} $\mathcal{C}$ decrypts $\left[x'_1\right]_c$ and $\left[x'_2\right]_c$. Clearly, $D_c(\left[x'_1\right]_c)$ is equal to $q \cdot n_I$ and $D_c(\left[x'_2\right]_c)$ is equal to $n_I$. $\mathcal{C}$ obtains $q$ after division.
\end{itemize}

\vspace{-3mm}
\subsection{Maximum Distance Query (MAXQ/C)}
\label{sec:maxdist3}

Let $q$ be the maximum distance between a user in $\mathcal{U_I}$ and her nearest facility. The protocol is defined as follows:
\begin{itemize}
\item{\textbf{Step 1:}} $\mathcal{C}$ sends the location of each facility to $\mathcal{S}$.
\item{\textbf{Step 2:}} As described in RNNQ/C protocol, the server aborts the protocol if it detects a threat. $\mathcal{S}$ calculates the distance between each facility and each user in $\mathcal{U_S}$. $\mathcal{S}$ determines the nearest facility of each user $U_i$ in $\mathcal{U_S}$ and the distance $d_i$ to the nearest facility. Let $max$ be the maximum distance between a user in $\mathcal{U_S}$ and her nearest facility. $\mathcal{S}$ selects a value $w$, which is greater than $max$.
\item{\textbf{Step 3:}} For each $j \in \left\{1,..,w\right\}$, $\mathcal{S}$ calculates the multiplication of $\left[T_i\right]_c$ values such that $U_i \in \mathcal{U_S}$ and $d_i = j$. That is, $\mathcal{S}$ computes $\left[x_j\right]_c = \prod_{U_i \in \mathcal{U_S} \& d_i = j} \left[T_i\right]_c$. If there is no such $U_i$, $\mathcal{S}$ sets $\left[x_j\right]_c = E_c(0)$. Therefore, $\left[x_j\right]_c$ is equal to the encryption of the total number of users in $\mathcal{U_I}$ whose distance to the nearest facility is equal to $j$. The query result $q$ is equal to the maximum $j$ value such that $D_c(\left[x_j\right]_c) \neq 0$.
\item{\textbf{Step 4:}} At the end of the protocol, $\mathcal{C}$ should not learn anything more than the query result. To hide the $\left[x_j\right]_c$ values from $\mathcal{C}$, $\mathcal{S}$ randomizes the $\left[x_j\right]_c$ values by exponentiation. $\mathcal{S}$ selects $w$ random values $\left\{v_1,...,v_w\right\}$. Then, $\mathcal{S}$ calculates $\left[x'_i\right]_c = \left[x_i\right]_c^{v_i}$ for each $i \in \left\{1,2,...,w\right\}$. If $\left[x_i\right]_c$ is the encryption of 0, $\left[x'_i\right]_c$ is the encryption of 0. Therefore, $q$ is still equal to the maximum $j$ value such that $D_c(\left[x'_j\right]_c) \neq 0$.
\item{\textbf{Step 5:}} $\mathcal{S}$ sends $\left[X'\right]_c = \left\{\left[x'_1\right]_c,...,\left[x'_w\right]_c\right\}$ to $\mathcal{C}$.
\item{\textbf{Step 6:}} $\mathcal{C}$ decrypts all $\left[x'_i\right]_c$ values. $\mathcal{C}$ obtains $q$, since it is equal to the maximum $j$ value such that $D_c(\left[x'_j\right]_c) \neq 0$.
\end{itemize}

\vspace{-3mm}
\subsection{Security Analysis of Client-Based Protocols}
\label{security-client}

In this section, we prove the security of the client-based protocols using the simulation paradigm described in Section \ref{sec:evalServ}. To prove the security of the protocols we need to show that there exists two probabilistic polynomial-time simulators $Sim_1$ and $Sim_2$ for simulating the views of the client and the server, respectively. In client-based protocols, the view of the client only consists of its input and output. The server only sends the encrypted query result to the client in Step 5. Since, the encrypted result is anonymized in Step 4, $\left[X'\right]_c$ does not contain any information about the users. Therefore, the view of the client can obviously be simulated by $Sim_1$ and the privacy of the server is assured.

The view of the server ($\textsc{VIEW}_2$) consists of $\mathcal{U_S}$, user locations, $\mathcal{F}$, and $\left[\mathcal{T}\right]_c$. To prove that the client's privacy is assured in the protocol, we need to show that there exists a probabilistic polynomial-time simulator $Sim_2$ such that $Sim_2(\mathcal{U_S},$ user locations$,~\mathcal{F}) \stackrel{c}{\equiv} \textsc{VIEW}_2$. This is satisfied by letting $Sim_2$ generate $n$ random numbers between $0$ and $m_c^2$. These numbers are computationally indistinguishable from the ciphertexts in $\left[\mathcal{T}\right]_c$ due to the semantic security of Paillier cryptosystem. Hence, we conclude that the client-based protocols privately process the queries in semi-honest model.

\vspace{-3mm}
\section{Protocols with Differential Privacy}
\label{difPriv}

Differential privacy is a framework to formalize privacy in statistical databases. The security proofs indicate that the proposed protocols reveal no more information than the output of the queries. However, providing aggregate statistical information about a database may reveal information about the individuals in the dataset. All of the queries (RNNQ, AVGQ, and MAXQ) that are studied in this paper return aggregate results and these query results may cause information leaks in some cases. For example, RNNQ returns the cardinality of $RNN(F_i)$ for each facility $F_i$ in $\mathcal{F}$. If the RNN cardinality of a facility is 1, this user can be predicted with background knowledge. However, only a region containing the user's location can be inferred. In any case, it is not possible to find the exact location of a user.

The protocols defined in Section \ref{sec:server} and Section \ref{sec:client} return exact query results. To achieve differential privacy in these protocols, we need to add controlled random noise to the query result. As discussed in Section \ref{sec:difPriv}, one needs to define the sensitivity of a query to determine the amount of noise to be added to the result of a query. Now, we show the sensitivity of each considered query and how to add the noise during the protocols.

\textbf{RNNQ} returns the total number of users attracted by each facility. It can be thought as a histogram query \cite{dwork2006calibrating} and its sensitivity is 2. When there is a single change in the database, RNN of at most two facilities may change. Therefore, we add a noise $Laplace(\frac{2}{\epsilon})$ to the RNN cardinality of each facility.

\textbf{AVGQ} returns two values: (i) the total number of users in $\mathcal{U_I}$ ($n_I$) and (ii) the total distance between each user and her nearest facility ($q \cdot n_I$). Thus, we need to calculate the sensitivity for both subqueries. Since the total number of users is a counting query, the sensitivity for $n_I$ is 1. For the total distance, the sensitivity is the maximum distance ($max$) between a user in $\mathcal{U_S}$ and her nearest facility. Therefore, we add the noise from $Laplace(\frac{1}{\epsilon})$ to $n_I$ and $Laplace(\frac{max}{\epsilon})$ to $q \cdot n_I$.

\textbf{MAXQ} returns $w$\footnote{$w$ is a random number that is selected by the server in the MAXQ/S and MAXQ/C protocols} values containing zero and non-zero elements. The largest index of a non-zero element is the result of the query. In MAXQ each of $w$ values can be considered as a counting query, and hence the sensitivity of each one is 1. Therefore, we add $Laplace(\frac{1}{\epsilon})$ to each $w$ values.

In the server-based protocols, the server adds noise to the query result in Step 8. Before sending the masked result $X''$ to the client, the server adds noise to the masked result. When the client applies unmasking in Step 10, it obtains the noisy result instead of the exact result.

In the client-based protocols, the server adds noise to the query result in Step 4. Before sending the encrypted result to the client, the server anonymizes the result by multiplying it with the encryption of zero in the client-based protocols. Instead of encrypting zero values, the server encrypts the values drawn from the Laplace distribution and multiplies the encryption of the noise with the encrypted query result. Due to homomorphic properties of Paillier cryptosystem, the noise will be added to the query result in plaintext. When the client decrypts query result in Step 6, it obtains the noisy result instead of the exact result.

\vspace{-3mm}
\section{Evaluation}
\label{eval}

In this section, we analyze the complexity, performance, and the utility of the proposed protocols. As there is no existing work that solves the stated problems, we only show the feasibility of our solutions. Firstly, we analyze the computation complexity and the communication costs theoretically in Section \ref{complex}. In Section \ref{effic}, we present the experimental efficiency evaluation of each protocol with respect to different parameters. In Section \ref{utility}, we show the utility of the protocols when differential privacy is achieved.

\vspace{-3mm}
\subsection{Complexity Analysis}
\label{complex}

In this section, we analyze the computation and communication costs of the proposed protocols in Section \ref{sec:server} and Section \ref{sec:client}. Achieving differential privacy as described in Section \ref{difPriv} does not change the communication costs of the protocols. Moreover, its effect on computation time is negligible because only overhead to achieve differential privacy is producing the noise drawn from the Laplace distribution. Therefore, we give the computation costs of the protocols as described in Section \ref{sec:server} and Section \ref{sec:client}.

\textbf{Server-based protocols.} Table~2 shows the total number of operations performed during server-based protocols in terms of total number of encryptions, decryptions, multiplications, exponentiations, distance calculations, and permutations. In all protocols, encryptions dominate the computation times. The number of encryptions is proportional to $n$ and the server performs at least $n$ encryptions in each query. However, the server encrypts 0 or 1 in each encryption and it can encrypt these values offline before the protocol. When the server uses precomputed $E_s(0)$ and $E_s(1)$ values in these protocols, computation cost reduces significantly. In addition, all of these ciphertexts must be transferred to the client in each query. Hence, the computation costs of RNNQ/S, AVGQ/S, and MAXQ/S are $n \cdot k$, $2 \cdot n$, and $n \cdot w$ ciphertexts, respectively.

\begin{table}
\centering
\caption{Computation performed in proposed protocols.}
\vspace{-2mm}
\begin{tabular}{|c|c|c|} \hline
& $\mathcal{S}$ & $\mathcal{C}$ \\ \hline
\textbf{RNNQ/S}& \begin{tabular}{c} $n_s \cdot k$ dist. \\ $n \cdot k$  enc. \\ $k$ dec. \end{tabular}
& \begin{tabular}{c} $n_c \cdot k$ mult. \\ $k$ enc. \end{tabular}   \\ \hline
\textbf{AVGQ/S}& \begin{tabular}{c} $n_s \cdot k$ dist. \\ $2 \cdot n$ enc. \\ $2$ dec. \end{tabular}
&\begin{tabular}{c} $2 \cdot n_c$ mult. \\ $2$ enc. \\ $1$ div. \end{tabular}  \\ \hline
\textbf{MAXQ/S}& \begin{tabular}{c} $n_s \cdot k$ dist. \\ $n \cdot w$ enc. \\ $w$ dec. \end{tabular}
&\begin{tabular}{c} $w \cdot (n_c - 1)$ mult. \\ $w$ exp. \\ $2$ per. \end{tabular} \\ \hline
\textbf{RNNQ/C}& \begin{tabular}{c} $n_s \cdot k$ dist. \\ $k$ enc. \\  $n_s + k$ mult. \end{tabular}
& \begin{tabular}{c} $k$ dec. \end{tabular}   \\ \hline
\textbf{AVGQ/C}& \begin{tabular}{c} $n_s \cdot k$ dist. \\ $2$ enc. \\ $2 \cdot n_s$ mult. \\ $n_s$ exp. \end{tabular}
&\begin{tabular}{c}  $2$ dec. \\ $1$ div. \end{tabular}  \\ \hline
\textbf{MAXQ/C}& \begin{tabular}{c} $n_s \cdot k$ dist. \\ $n_s$ mult. \\ $w$ exp. \\ $\leq w$ enc. \end{tabular}
& \begin{tabular}{c} $w - q + 1$ dec. \end{tabular} \\
\hline\end{tabular}
\vspace{-3mm}
\end{table}

\textbf{Client-based protocols.} In the setup phase of the client-based protocols, $n$ encryptions are computed by the client. The client sends these $n$ ciphertexts to the server in the setup. Therefore, the communication overhead of the setup is $n$ ciphertexts. After completion of the setup phase, all queries can be processed with small computation and communication overheads. Table 2 shows computation costs of client-based protocols in each query. Total number of encryptions in each query is very small with respect to the server-based protocols. The computation costs of RNNQ/C, AVGQ/C, and MAXQ/C are $k$, $2$, and $w$ ciphertexts, respectively.

\vspace{-3mm}
\subsection{Efficiency}
\label{effic}

We have implemented the protocols in Java and we used the implementation in \cite{paillierImp} for Paillier cryposystem. All experiments were performed on a 64-bit Windows 7 machine with 2.6 GHz Intel Core i5 processor and 4 GB of RAM. We used 1024-bit modulus $m_s$ and $m_c$ in our tests and each ciphertext consists of 2048 bits. All distances were calculated by the server in the Euclidean metric. 

In our experiments, we used real datasets \cite{yang2014modeling} containing 227,428 check-ins in New York City and 573,703 check-ins in Tokyo. The $x$ and $y$ coordinates were scaled to integer values from 1 to 10,000. Since the total number of users in the datasets is less than 5,000, we considered each check-in location as the location of a separate user. Therefore, $n_s = 227,428$ in NYC dataset and $n_s = 573,703$ in Tokyo dataset. We randomly chose $20\%$ of them as the users of the client. For existing facilities, we used the locations of 20 restaurants of a fast food chain in New York and 10 restaurants of a fast food chain in Tokyo.

For synthetic datasets, the $x$ and $y$ coordinates of the user locations and facility locations were selected randomly as integer values from 1 to $maxCoordinate$. The user id values in the superset $\mathcal{U}$ were selected as the numbers from 1 to $n$. We randomly chose $n_s$ of them as the users of the server and $n_c$ of them as the users of the client. The key parameters in the implementation were $n$, $n_s$, $n_c$, $n_I$, $k$, $maxCoordinate$, $m_s$, and $m_c$ (introduced in Table~1). $w$ is another parameter in the Maximum Distance Query, which depends on the value of $maxCoordinate$. We present the experimental evaluation of the server-based protocols in Section~\ref{efficServer} and the client-based protocols in Section~\ref{efficClient}.

\vspace{-2mm}
\subsubsection{Server-based Protocols}
\label{efficServer}

When we use 1024-bit $m_s$ in Paillier encryption, one million encryption nearly takes 2 hours and 45 minutes and the size of one million ciphertexts is 250 MB. For the protocols RNNQ/S, AVGQ/S, and MAXQ/S, the computation times and communication costs are directly proportional to $n \cdot k$, $2 \cdot n$, and $n \cdot w$, respectively. Therefore, when $n$ is one million, the computation time of each protocol is more than 2 hours and 45 minutes. Moreover, when $n$ is one million, the amount of data exchanged during each protocol is more than 250 MB.

In our experiments, we set $n = 1,000,000$, $n_s = 100,000$, $n_c= 20,000$, $k = 25$, and $maxCoordinate = 10,000$ in the synthetic dataset. Running time of RNNQ/S with these parameters is high because it requires $n \cdot k$ encryptions for the encrypted matrix $\left[\mathcal{T}\right]_s$. However, all of these ciphertexts in $\left[\mathcal{T}\right]_s$ are either the encryption of 0 or the encryption of 1. Therefore, the encrypted values in $\left[\mathcal{T}\right]_s$ can be computed offline by the server. When the server precomputes $E_s(0)$ and $E_s(1)$ values before the protocol, the remaining computation takes 10 seconds for these parameters. For the NYC and Tokyo datasets, the query takes 20 seconds and 25 seconds, respectively. Similarly, for the synthetic dateset with given parameters, AVGQ/S takes 13.5 seconds, when the server computes $E_s(0)$ and $E_s(1)$ values before the protocol. Since the computation time of AVGQ/S is directly proportional to $n_s$, the query takes nearly 35 and 70 seconds for the NYC and Tokyo datasets, respectively. The computation time of MAXQ/S mostly depends on the value of $w$, which is a randomly selected number by the server. The server performs $w$ decryptions and the client performs $w$ exponentiations and $w \cdot (n_c - 1)$ multiplications. For instance, when $w$ is selected as 500, the computation time of MAXQ/S is nearly 5 minutes and it increases linearly when $w$ increases. When $w$ remains same, we observed the similar computation times for the real datasets.

Our experimental results show that the computation at the client's side is low in server-based query processing protocols. Step 3 of these protocols necessitates calculating an encrypted matrix $\left[\mathcal{T}\right]_s$. This step dominates the computation time of server-based protocols. However, the encrypted values can be computed offline by the server. To do offline computation, the server does not need to know the facility locations. The server can compute $E_s(0)$ and $E_s(1)$ values before the protocol. When the client sends the facility locations in a protocol, the server uses previously computed ciphertexts in the encrypted matrix $\left[\mathcal{T}\right]_s$. Hence, if the server computes these encryptions offline before the protocol, the remaining computation takes a few minutes on a single computer for millions of users. In addition, the computation time of calculating $\left[\mathcal{T}\right]_s$ can be reduced via parallel computations because all encryptions are independent. Server-based protocols can be preferable when the client cannot afford to perform the computations or when the client wants to outsource all the computations to the server. However, as we have shown, the encrypted values in $\left[\mathcal{T}\right]_s$ must be transferred to the client in each query.

\vspace{-2mm}
\subsubsection{Client-based Protocols}
\label{efficClient}

In Section \ref{complex}, the computation complexity of each client-based protocol is given. When the client performs several queries, some of these computations are common. For instance, the server calculates the distance between each facility and each user in each query. For the same facilities in separate queries, the server does not need to calculate the same distance values. In our system model, the locations of the existing facilities are considered as public and known by the server. The client can share these locations in the setup phase. Since the server knows the locations of the existing facilities, we assume that all the distances between users and existing facilities were calculated and the nearest facility of each user was determined by the server before the execution of protocols. In each query, the client sends a possible location for adding a new facility and the server only calculates the distance between the new location and each user. The server only updates the nearest facilities of the users who are attracted by the new facility. Therefore, we evaluate the following for each protocol under different parameter settings:
\begin{itemize}
	\item \textbf{Precomputation time.} Most of the computation given in Table 2 can be precomputed by the server because the locations of the existing facilities are known by the server. Hence, we evaluate the precomputation time of each protocol separately.
	\item \textbf{Query processing time.} Once the server completes the precomputation, processing of each query requires low computation overhead. We evaluate the query processing time when the client sends a possible location for adding a new facility.
	\item \textbf{Amortized computation time.} When the client requests $n_q$ queries, amortized computation time of a query is equal to ((precomputation time) + $n_q \cdot $ (query processing time)) / $n_q$.
\end{itemize}

In the setup phase of the client-based protocols, the client computes $n$ ciphertexts and shares them with the server. Therefore, the computation cost and the communication cost of the setup phase of the client-based protocols are directly proportional to $n$. One ciphertext consists of 2048 bits and one encryption takes 10 ms on the machine mentioned above. Therefore, if $n$ is one million, the execution time of the setup phase is nearly 2 hours and 45 minutes\footnote{Computation time can be further reduced via parallel computations.} and the amount of data sent by the client to the server is 250 MB.

Table 2 shows computation costs of client-based protocols including precomputation and query processing. We evaluate the performance of the protocols with respect to $n$, $n_s$, $n_c$, $n_I$, $k$, and $maxCoordinate$. As evident in Table 2, the parameters $n$, $n_c$, and $n_I$ have no effect on query processing times of the protocols. In our experiments, we observed the similar computation times for different values of these parameters. Therefore, increasing one of these parameters does not change the precomputation time and the query processing time of client-based protocols. For the other parameters $n_s$, $k$, and $maxCoordinate$, we analyze their effects on the precomputation time, the query processing time and the communication cost of each client-based protocol. In our experiments, we set $n = 200,000,000$, $n_s = 5,000,000$, $n_c= 1,000,000$, $n_I = 500,000$, $k = 100$, and $maxCoordinate = 10,000$, unless stated otherwise.

\begin{figure*}[t!]%
\centering
  \includegraphics[height=1.45in]{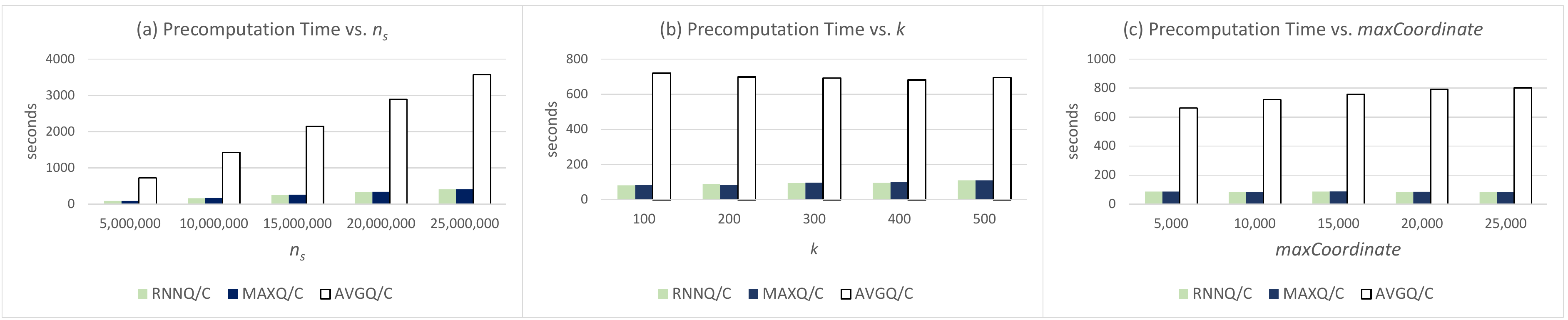}
		\caption{Precomputation times of client-based protocols.}
		\vspace{-2mm}
\label{PreClient}%
\end{figure*}

\begin{figure*}[t!]%
\centering
  \includegraphics[height=1.45in]{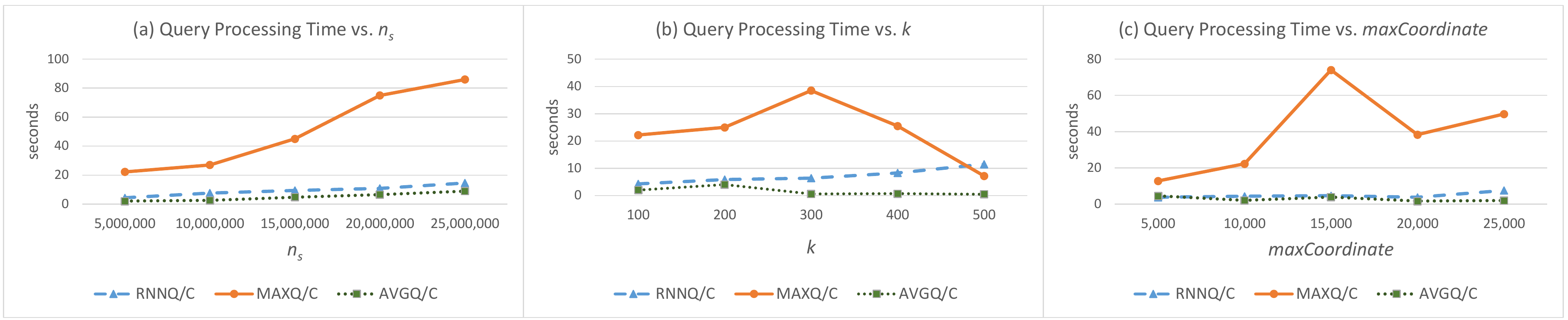}
		\caption{Query processing times of client-based protocols.}
		\vspace{-3mm}
\label{QpcClient}%
\end{figure*}

\textbf{RNNQ/C.} Table 2 shows the computation cost of the protocol, where $n_s$ and $k$ are the determining parameters. In this protocol, $n_s \cdot k$ distance calculations, $n_s$ multiplications and $k$ encryptions can be precomputed by the server. The client computes $\left[X\right]_c = \left\{\left[x_1\right]_c,...,\left[x_k\right]_c\right\}$ before the protocol. When the client requests a query for a new facility location ($F_{k+1}$), the server only calculates the distance between $F_{k+1}$ and each user. Then, the client multiplies the $\left[T_i\right]_c$ values of the users whose nearest facility is $F_{k+1}$ and calculates $\left[x_{k+1}\right]_c$. The client also multiplies the inverse of $\left[T_i\right]_c$ values of the same users with the $x_j$ values of their previous nearest neighbors. Therefore, during query processing the server performs $n_s$ distance calculations, nearly $2 \cdot \frac{n_s}{k}$ multiplications, and nearly $\frac{n_s}{k}$ modular inverse calculations. The client also performs $k$ decryptions during query processing.

Although encryption and decryption are more expensive operations than multiplication, the most time consuming part in the precomputation time is $n_s$ multiplications because $n_s$ is much higher than $k$ in our experiments. Therefore, the precomputation time mostly depends on $n_s$ and slightly depends on $k$. Figure 5(a) illustrates the effect of $n_s$ on precomputation time. The precomputation time increases from 82 seconds to 407 seconds, when $n_s$ increases from 5 million to 25 million. As evident in Figure 5(b), the effect of $k$ is not sharp as $n_s$. For instance, the precomputation takes 82 seconds when $k$ is 100. When $k$ becomes 500, the time increases to 110 seconds. For the NYC and Tokyo datasets, the precomputation time is 4 seconds and 9.6 seconds, respectively, due to the lower $n_s$ values in these datasets.

The query processing time of RNNQ/C depends on the values of $n_s$ and $k$. Although an increase in $k$ decreases the workload of the server, it increases the total number of decryptions performed by the client. Figure 6(a) and Figure 6(b) shows the query processing time for different values of $n_s$ and $k$. These two variables are not the only factors that determine the query processing time because the total number of operations also depends on the total number of users attracted by the new facility. Query processing takes 1.9 seconds and 2.5 seconds, for the NYC and Tokyo datasets, respectively.

We also evaluate the amortized computation time of RNNQ/C for 100 queries. For the parameters given above, the precomputation takes 82 seconds and the query processing takes 4.3 seconds. Hence, the amortized computation time per query is 5.1 seconds.

During query processing, $k$ ciphertexts and the facility locations are shared between the server and the client. Hence, $k$ is the most crucial parameter for the communication cost. When the total number of facilities is 100, the amount of shared data is nearly 25 KB.

\textbf{AVGQ/C.} In this protocol, the client obtains the total number of users in $\mathcal{U_I}$ and the total distance between each user and her nearest facility. Until there is a change on the total number of users, there is no need to compute it in each query. Hence, the total number of users in $\mathcal{U_I}$ can be precomputed by the server and the client. This part of precomputation requires $n_s$ multiplications, one encryption, and one decryption. In addition, the server can precompute $n_s \cdot k$ distance calculations, $n_s$ exponentiations, $n_s$ multiplications, and one encryption for the computation of the total distance. During query processing the server performs $n_s$ distance calculations, nearly $2 \cdot \frac{n_s}{k}$ multiplications, nearly $\frac{n_s}{k}$ exponentiations, and nearly $\frac{n_s}{k}$ modular inverse calculations. The client performs one decryption and one division during query processing.

Due to the high number of exponentiations, the precomputation time of AVGQ/C is higher than the other client-based protocols. Figure 5(a) depicts the precomputation time for different values of $n_s$. Query processing takes nearly 12 minutes when the server has 5 million users and the time changes linearly with respect to the value of $n_s$. For the smaller $n_s$ values in NYC and Tokyo datasets, the precomputation takes 41 seconds and 92 seconds, respectively. In addition, Figure 5(c) shows that the value of $maxCoordinate$ affects the precomputation time slightly. As $maxCoordinate$ increases, exponent values in the computation also increase.

The query processing time of AVGQ/C is directly proportional to $n_s$ and inversely proportional to $k$. Figure 6(a) and Figure 6(b) shows the query processing time for different values of $n_s$ and $k$. Since there is no encryption and only one decryption in query processing, AVGQ/C has the lowest query processing time among three client-based protocols. Query processing takes nearly 1 second, for both NYC and Tokyo datasets. Similar to RNNQ/C, the total number of users attracted by the new facility affects the query processing time. The precomputation takes 720 seconds and the query processing takes 2 seconds for the parameters given above. Hence, the amortized computation time per query is 9.2 seconds for 100 AVGQ/C queries.

During the protocol, two ciphertexts and the facility locations are shared between the server and the client. Therefore, the communication cost is low because the facility locations are sent as plaintext and the total number of ciphertexts is two. When the total number of facilities is 100, the amount of shared data is less than 1 KB during query processing.

\textbf{MAXQ/C.} In MAXQ/C, the server can precompute $n_s \cdot k$ distance calculations, $n_s$ multiplications and $w$ encryptions. During query processing, the server performs $n_s$ distance calculations, nearly $2 \cdot \frac{n_s}{k}$ multiplications, nearly $\frac{n_s}{k}$ modular inverse calculations, and $w$ encryptions. The client also performs $w - q + 1$ decryptions during query processing.

$w$ and $n_s$ are crucial parameters in the precomputation time of MAXQ/C. Similar to RNNQ/C protocol, the precomputation time mostly depends on $n_s$ because $n_s$ is much higher than $w$ in our experiments. Figure 5(a) shows the precomputation time with respect to $n_s$. Since $n_s$ multiplications dominate the computation cost, the precomputation time of MAXQ/C is similar to RNNQ/C. We also observed similar results for the real datasets.

Due to the randomness in the selection of $w$, the query processing time of MAXQ/C is not directly proportional to $n_s$ or $maxCoordinate$. $w$ is a randomly selected value that is greater than $max$, which is the maximum distance between a user in $\mathcal{U_S}$ and her nearest facility. In our experiments, $w$ is selected randomly in the range [$max$, $2 \cdot max$]. Therefore, the parameter $maxCoordinate$ affects the value of $w$, and hence the query processing time. The query processing time of MAXQ/C for different values of $maxCoordinate$ is given in Figure 6(c). As evident in Figure 6(c), query processing time is not directly proportional to $maxCoordinate$. For instance, when $maxCoordinate$ increases from 15,000 to 20,000, the computation time decreases due to a decrease in $w$ and an increase in $q$. Therefore, the distance of each user to her nearest facility and the query result $q$ also affect the query processing time. In addition, Figure 6(b) shows the query processing time for different values of $k$. Smaller $k$ values may result in higher query processing times because $max$ value may increase in smaller $k$ values. For instance, in real datasets we have smaller $k$ values such as 10 and 20. As a result, the query processing times are 17 seconds and 60 seconds, for the NYC and Tokyo datasets, respectively.

For the parameters given above, the precomputation takes 83 seconds and the query processing takes 22.2 seconds. Therefore, the amortized computation time per query is 23 seconds for 100 MAXQ/C queries. The communication cost of the protocol is $w$ ciphertexts and the facility locations. When $w$ is selected as 5000, the amount of shared data is nearly 1.25 MB.

All these results show the practicality of our proposed scheme in real-life settings. Any of the aforementioned queries can be performed in less than a minute on datasets that include millions of individuals.

\vspace{-3mm}
\subsection{Utility vs. Differential Privacy}
\label{utility}

In Section \ref{difPriv}, we explain how to achieve differential privacy in the proposed protocols by adding controlled noise to the query results, which affects the accuracy of the results. To measure the utility of the protocols under differential privacy, we selected 100 candidate locations for the new facility and observed the results after executing the protocols. We divided the whole region into a 10x10 grid and selected the center of each grid as a candidate location for the new facility. First, we applied the protocols without adding any noise and ranked the 100 candidate locations with respect to their optimality. Then, we executed the protocols by adding controlled noise and observed the impact of differential privacy on the utility. We evaluated the utility of differential privacy for the real and synthetic datasets. For synthetic datasets, we set the parameters as given in Section \ref{effic}.

\begin{figure*}[t!]%
\centering
  \includegraphics[height=1.45in]{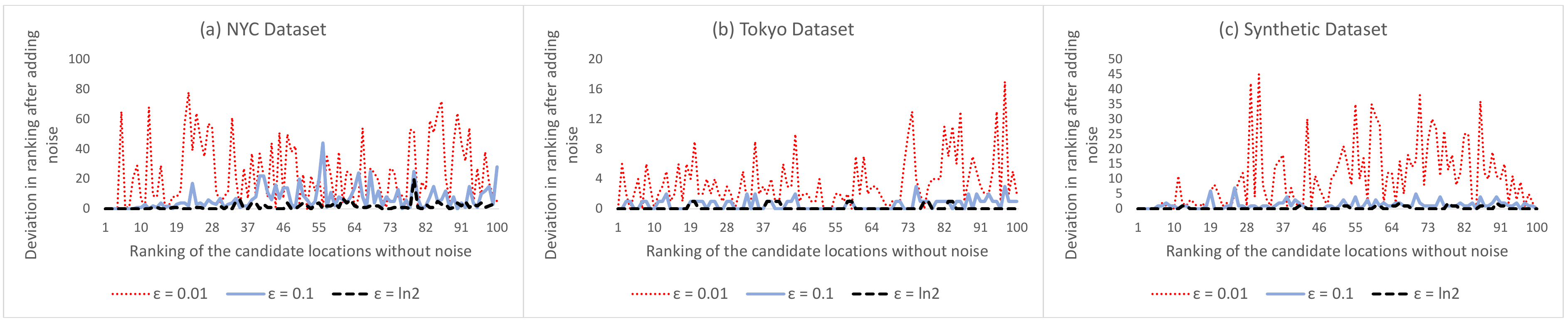}
		\caption{Deviation in the rankings of the 100 candidate locations after achieving differential privacy in RNNQ. $x$ axis represents the ranking of the candidate locations when RNNQ is performed with exact query results. $y$ axis represents the change in the ranking of each candidate location when differential privacy is achieved in RNNQ. For instance, in Figure 7(a) the point $(5,65)$ for $\epsilon = 0.01$ shows that the 5th best candidate location for the new facility becomes 70th best location after adding noise to the query result. Depending on the maximum change in the ranking, the range of $y$ axis varies for each dataset.}
		\vspace{-2mm}
\label{utility1}%
\end{figure*}

\begin{figure*}[t!]%
\centering
  \includegraphics[height=1.45in]{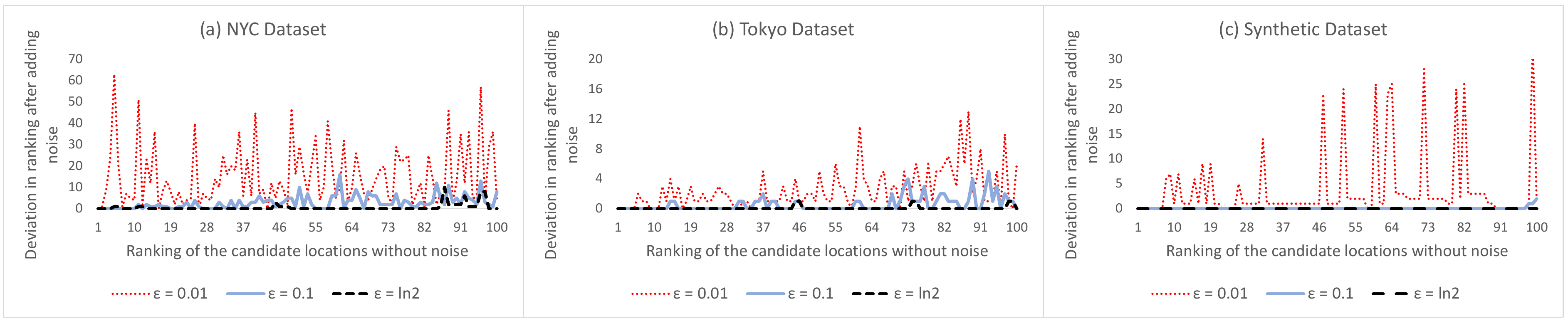}
		\caption{Deviation in the rankings of the 100 candidate locations after achieving differential privacy in AVGQ.}
		\vspace{-3mm}
\label{utility2}%
\end{figure*}

\textbf{RNNQ.} The objective of using RNNQ is uniformly distributing the cardinality of the RNNs. When the new facility attracts users from dense facilities, the workloads of dense facilities decrease. Hence, balancing workload reduces the wait times by avoiding overloads. We measured the standard deviation of the cardinalities of the RNNs. We sorted 100 candidate locations with respect to the standard deviation after adding the possible location as the new facility. The best candidate is the location that minimizes the standard deviation. We also sorted the candidate locations after achieving differential privacy. Figure 7 shows the rankings of the candidates for real and synthetic datasets after adding controlled noise. We used three different $\epsilon$ values such as 0.01, 0.1, and $\ln2$, which are typically chosen values in the literature. As evident in Figure 7, the utility increases when $\epsilon$ increases. When $\epsilon$ is $\ln2$, the ranking of the candidates are almost same as the rankings without adding any noise. Although the deviations in the rankings increase for the smaller values of $\epsilon$, the best candidate is same most of the time after achieving differential privacy. Therefore, the utility of RNNQ under differential privacy is remarkable for large $\epsilon$ values and acceptable for small $\epsilon$ values.

\textbf{AVGQ.} We sorted 100 candidate locations with respect to the average distance value returned from the query. The best candidate is the location that minimizes the average distance. Figure 8 shows the rankings of the candidates after adding controlled noise. The deviation on the rankings is less than RNNQ. Therefore, the utility of AVGQ under differential privacy is better than RNNQ for all $\epsilon$ values.

\textbf{MAXQ.} This query returns $w$ values containing zero and non-zero elements. The largest index of a non-zero element is the result of the query. After adding the noise to each of $w$ values, most of the zero values becomes non-zero. Therefore, adding the noise changes the query result significantly. In our experiments, we observed that the query result becomes $w$ or $w-1$ most of the time after adding the noise. Since $w$ is a randomly selected value, the query returns a random result in each execution. Hence, the utility of MAXQ under differential privacy is very low.

\textbf{Discussion.} Our experimental results show that achieving differential privacy in RNNQ and AVGQ causes low utility loss. Since these queries contain counting subqueries, running them under differential privacy increases the privacy of individuals with a negligible computational overhead. On the other hand, MAXQ is not suitable for differential privacy because the query result changes significantly after adding noise to each counter value in the query. To prevent high utility loss of differential privacy in MAXQ, only non-zero values should be randomized as described in Section \ref{sec:maxdist3}.

\vspace{-3mm}
\section{Conclusion}
\label{conc}

We proposed novel protocols for privacy-preserving analysis of location data in a location-based service provider (referred as the server) by a business (referred as the client) as a service. We defined three queries addressing different objectives in optimal location selection: (i) to minimize the average distance between each user and her closest facility, (ii) to minimize the maximum distance between a user and her closest facility, and (iii) to uniformly distribute the workload in facilities. We developed two homomorphic encryption-based solutions: (i) a server-based solution, in which most of the computation is performed by the server, and hence the workload of the client is low, and (ii) a client-based solution, in which the client performs the majority of the computation during the setup phase (which only occurs once) and after completion of the setup phase, all queries are processed quickly. We showed that the proposed protocols keep the client's user list and the query result hidden from the server, and the location information stored at the server hidden from the client. The security provided by all protocols relies on the underlying security of the Paillier cryptosystem (which relies on the decisional composite residuosity assumption) proved in \cite{paillier1999public}. We also showed that it is possible to achieve differential privacy in the proposed protocols with low utility loss. Using the proposed protocols will facilitate sharing location information between entities without compromising customer privacy. We evaluated the efficiencies of the proposed protocols through experiments for each considered query type and showed that the proposed protocols are feasible, efficient, and scalable.




\vspace{-10mm}
\begin{IEEEbiography}[\vspace{-7mm}{\includegraphics[width=1in,height=1.25in,clip,keepaspectratio]{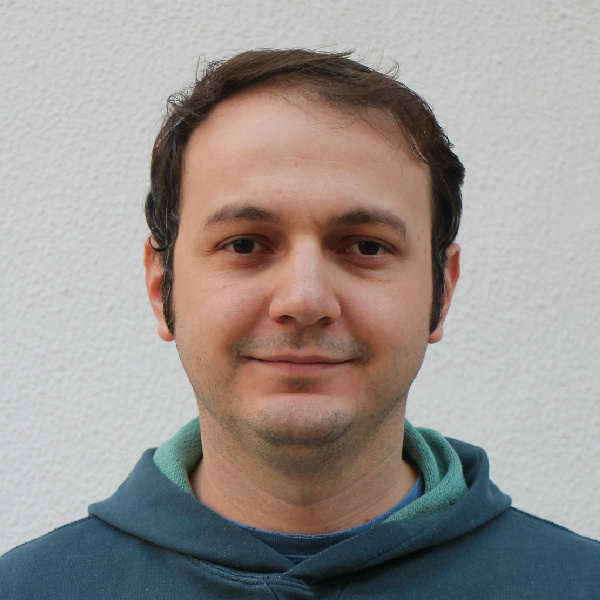}}]{Emre Yilmaz}
received the BS degree in computer science and engineering from Sabanci University, Istanbul, Turkey, in 2008 and the MS degree in computer science from ETH Zurich, Switzerland, in 2010. He is continuing working toward the PhD degree in computer engineering at Bilkent University, Ankara, Turkey. His research interests include data privacy, cryptography, and big data analytics.
\end{IEEEbiography}

\vspace{-16mm}
\begin{IEEEbiography}[{\includegraphics[width=1in,height=1.25in,clip,keepaspectratio]{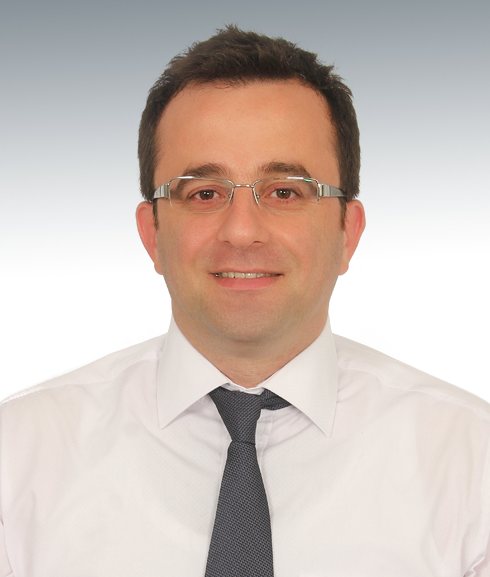}}]{Hakan Ferhatosmanoglu} received the PhD degree in computer science from the University of California Santa Barbara, in 2001. He is a professor with the University of Warwick, UK. His research is on scalable data management and analytics for multi-dimensional data. He received Research Awards from the US Department of Energy, US National Science Foundation, The Academy of Science of Turkey, and Alexander von Humboldt Foundation.
\end{IEEEbiography}

\vspace{-11mm}
\begin{IEEEbiography}[{\includegraphics[width=1in,height=1.25in,clip,keepaspectratio]{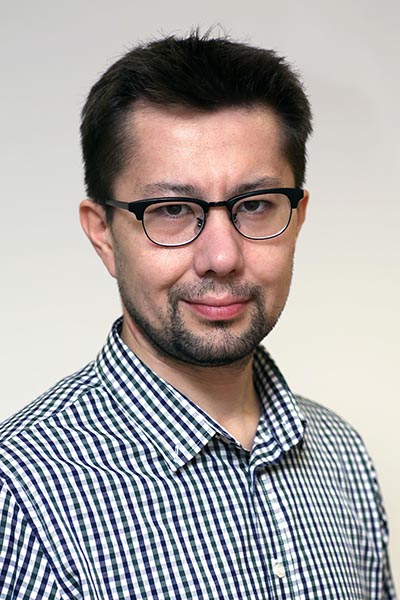}}]{Erman Ayday} received the MS and PhD degrees from the Georgia Institute of Technology, in 2007 and 2011, respectively. He is an assistant professor with Bilkent University, Turkey. Before that, he was a Post-Doctoral Researcher with \'Ecole Polytechnique F\'ed\'erale de Lausanne, Switzerland. His research interests include privacy-enhancing technologies (including big data and genomic privacy), wireless network security, trust and reputation management, and recommender systems.
\end{IEEEbiography}

\vspace{-10mm}
\begin{IEEEbiography}[\vspace{-7mm}{\includegraphics[width=1in,height=1.25in,clip,keepaspectratio]{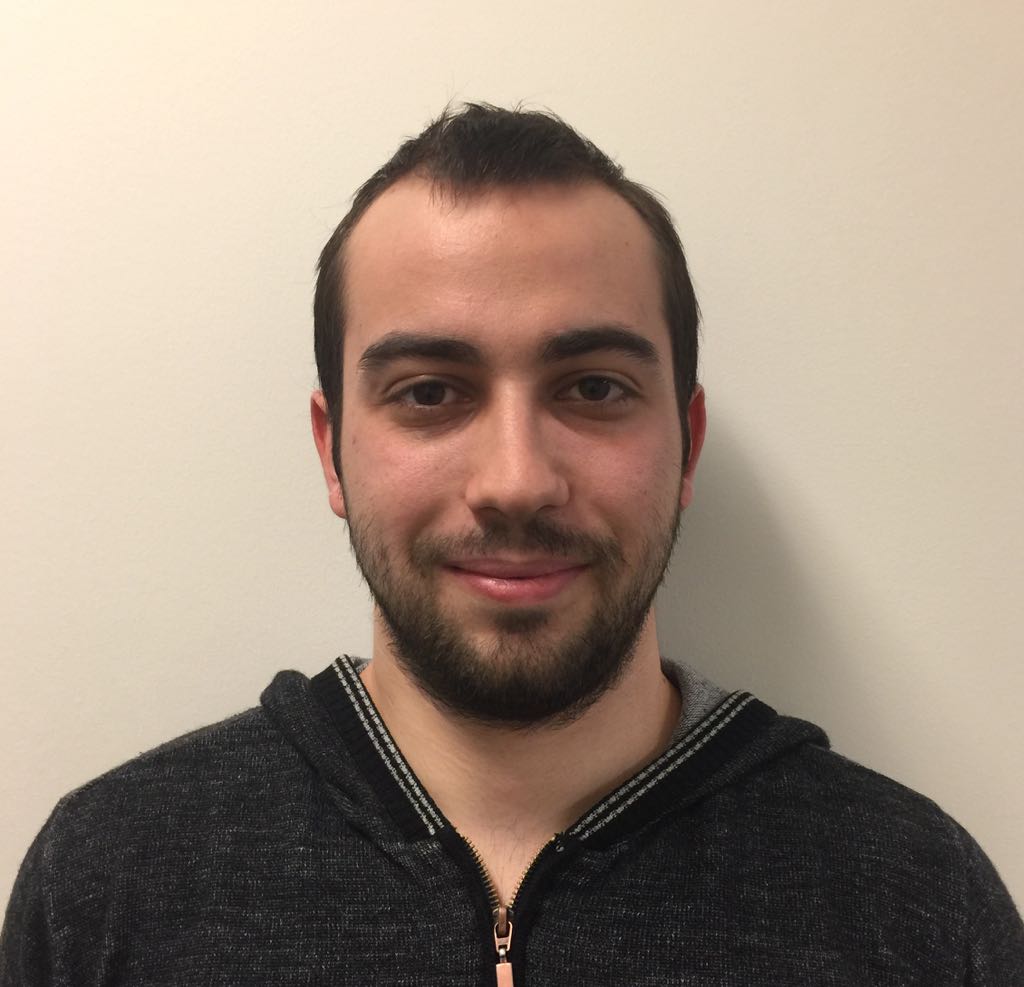}}]{Remzi Can Aksoy} received the BS degree in computer engineering from Bilkent University, Turkey, in 2016. He started the graduate studies in computer science at University of Michigan, Ann Arbor. His research interests include data privacy and big data.
\end{IEEEbiography}

\end{document}